\documentclass[aps,prd,twocolumn,superscriptaddress,showpacs,floatfix]{revtex4}

\usepackage{amssymb,amsmath}
\usepackage{graphicx}
\usepackage{dcolumn}
\usepackage{myscriptmath}

\newcommand{\cc}[1]{\multicolumn{1}{c}{#1}}

\newcommand{\nn}{\nonumber}
\newcommand{\D}{\displaystyle}
\newcommand{\T}{\textstyle}
\renewcommand{\S}{\scriptstyle}
\renewcommand{\SS}{\scriptscriptstyle}
\newcommand{\eg}{\textit{e.g.}}
\newcommand{\cf}{\textit{cf.}}
\newcommand{\ie}{\textit{i.e.}}

\newcommand{\etal}{\textit{et al.}}
\newcommand{\txl}{T$\chi$L}

\begin{document}

\preprint{WUB 05-03}
\preprint{FZJ-ZAM-IB-2005-01}

\title{Finite-Size Effects in Lattice QCD with Dynamical Wilson Fermions}


\author{Boris Orth}
\email[]{b.orth@fz-juelich.de}
\author{Thomas Lippert}
\affiliation{Bergische Universit\"at Wuppertal, Gau{\ss}stra{\ss}e 20,
  D-42097 Wuppertal, Germany}
\affiliation{John von Neumann Institute for Computing,
  Forschungszentrum J\"ulich, D-52425 J\"ulich, Germany}
\author{Klaus Schilling}
\affiliation{Bergische Universit\"at Wuppertal, Gau{\ss}stra{\ss}e 20,
  D-42097 Wuppertal, Germany}


\date{\today}

\begin{abstract}
As computing resources are limited, choosing the parameters for a full Lattice
QCD simulation always amounts to a compromise between the competing objectives
of a lattice spacing as small, quarks as light, and a volume as large as
possible. Aiming to push unquenched simulations with the Wilson action towards
the computationally expensive regime of small quark masses we address the
question whether one can possibly save computing time by extrapolating results
from small lattices to the infinite volume, prior to the usual chiral and
continuum extrapolations. In the present work the systematic volume dependence
of simulated pion and nucleon masses is investigated and compared with a
long-standing analytic formula by L\"uscher and with results from Chiral
Perturbation Theory (ChPT). We analyze data from Hybrid Monte Carlo
simulations with the standard (unimproved) two-flavor Wilson action at two
different lattice spacings of $a\approx$ 0.08~fm and 0.13~fm. The quark
masses considered correspond to approximately 85 and 50\% (at the smaller $a$)
and 36\% (at the larger $a$) of the strange quark mass. At each quark mass we
study at least three different lattices with $L/a=$10 to 24 sites in the
spatial directions ($L=$0.85--2.08~fm).
\end{abstract}


\maketitle

\section{\label{sec:intro}Introduction}

It is in the nature of any numerical Lattice QCD calculation that it can only
be done at non-zero lattice spacing and in finite volume.  Moreover, due to
limited computing resources the typical quark masses currently employed are
still substantially larger than the masses of the physical quarks. In order to
obtain physically meaningful predictions, extrapolations of lattice results to
the continuum, the infinite volume and to small quark masses are necessary.
In the context of spectrum calculations one usually extrapolates in the
lattice spacing and the quark mass, while the volume is preferably chosen such
that its systematic effect on the masses can be largely neglected. The
underlying assumption is that if the linear spatial extent $L$ of a lattice
with periodic boundary conditions is much larger than the Compton wavelength
of the pion (\eg\ if $\mpi L \gg 5$, according to a common rule of thumb),
then a single hadron $H$ is practically unaffected by the finite volume
(except that its momentum must be an integer multiple of $2 \pi/L$). Its mass
$m_H(L)$ in particular will be close to the infinite-volume value defined at
fixed lattice spacing and quark mass as
\begin{equation}
m_H \equiv \lim_{L\to\infty} m_H(L).
\end{equation}
If the box size is decreased until the hadron barely fits into the box, the
virtual pion cloud that surrounds the particle due to vacuum polarization is
distorted, and pions may be exchanged ``around the world''. As a consequence
the mass of the hadron receives corrections of order $e^{-\mpi L}$ to its
asymptotic value, which are small compared to the typical statistical errors
as long as the lattice remains sufficiently large. When $L$ gets very close to
the size of the region in which the valence quarks are confined, however, the
the quark wave functions of the enclosed hadron are distorted and one observes
rapidly increasing finite volume effects approximately proportional to some
negative power of $L$.

In the present work we explore the practical implications of this
picture by investigating, for various fixed values of the gauge
coupling and the quark mass, the actual volume dependence of simulated
light hadron masses. Against the background of our GRAL
project---whose name is an acronym for ``Going Realistic And
Light''---we ask in particular under which circumstances
extrapolations in the lattice volume could be appropriate to obtain
infinite-volume results from sub-asymptotic lattices, which would allow
one to save valuable computing time. To this end we compare our data
to various finite-size mass shift formulae available from the
literature.

While in past years the chiral extrapolation and the reduction of
discretization errors have been at the center of many theoretical and
numerical studies, there have, until recently, been rather few systematic
investigations into the lattice size dependence of light hadron masses. These
include, first of all, an analytic work of 1986 by L\"uscher
\cite{Luscher:1985dn} in which a universal formula for the asymptotic volume
dependence of stable particle masses in arbitrary massive quantum field
theories is proven. Some years later Fukugita \etal\
\cite{Fukugita:1991hw,Fukugita:1992jj,Fukugita:1992hr,Aoki:1993fq,Aoki:1993gi}
carried out a systematic investigation of finite-size effects in pion, rho and
nucleon masses from quenched and unquenched simulations (with the staggered
action). Related numerical studies with staggered quarks came also from the
MILC collaboration
\cite{Bernard:1993hv,Bernard:1993an,Gottlieb:1996hy}. Recently the systematic
dependence of light hadron masses and decay constants on the lattice volume
has been receiving renewed attention, see \eg\
Refs.~\cite{Becirevic:2003wk,Colangelo:2003hf,AliKhan:2003cu,Koma:2004wz,%
Detmold:2004ap,Guagnelli:2004ww,Arndt:2004bg,Beane:2004tw,Colangelo:2004xr,%
Bedaque:2004dt,Borasoy:2004zf,Thomas:2005qm,Colangelo:2005gd}. These studies
include, on the one hand, a determination of the pion mass shift in finite
volume using L\"uscher's asymptotic formula with input from infinite-volume
ChPT up to NNLO \cite{Colangelo:2003hf}. On the other hand, the finite-size
mass shift of the nucleon has been calculated using relativistic Baryon ChPT
in finite volume up to NNLO \cite{AliKhan:2003cu}. In the following
Section~\ref{sec:fse_formulae} we will briefly summarize those results to
which we will compare our numerical data in Section~\ref{sec:vol_dep}. The
details of the underlying simulations and the determination of light hadron
masses and other observables will be described in
Sections~\ref{sec:sim_details} and \ref{sec:observables}.

\section{\label{sec:fse_formulae}Finite-size mass shift formulae}

We consider a stable hadron $H$ ($=\pi,N$) on a four-dimensional
hypercubic space-time lattice of spatial volume $L^3$ and sufficiently
large time extent $T$, with lattice spacing $a$ set equal to unity for
convenience. Both the bare coupling $g$ and the quark mass held fixed,
for large $L$ the mass $m_H(L)$ of the hadron is supposed to become a
universal function of the product $m_\pi L$ in the finite-volume
continuum limit (which is obtained by taking $g\to 0$ and
simultaneously $L\to\infty$, while keeping $m_\pi L$ constant). Since
finite-size effects probe the system at large distances $L$ they are
insensitive to short-distance effects, so that this function should be
independent of the form and magnitude of any ultraviolet cut-off. It
is therefore expected to hold also for finite lattice spacings.

Attributing finite-volume effects at large $L$ to vacuum polarization
effects, L\"uscher's formula \cite{Luscher:1985dn} applied to QCD
relates the asymptotic mass shift
\begin{equation}
\Delta m_H(L) \equiv m_H(L)-m_H
\end{equation}
to the (infinite-volume) elastic forward scattering amplitude $F_{\pi
H}(\nu)$, where $\nu$ is the crossing variable. For the \emph{pion} it
is given in terms of $\Fpipi$ by \cite{Luscher:1983rk}
\begin{eqnarray}
\label{eqn:luescher_pion}
\Delta\mpi(L) &=& -\frac{3}{16\pi^2\mpi L} \opintlim{-\infty}{\infty}{dy}
e^{-\sqrt{\mpi^2+y^2}L}\Fpipi(iy) \nn \\ && +\,O(e^{-\bar{m}L}).
\end{eqnarray}
Because of $\bar{m}\ge\sqrt{3/2}\,\mpi$ the error term is
exponentially suppressed compared to the first term. Due to the
negative intrinsic parity of the pion and parity conservation in QCD
there is no 3-pion vertex, so that the term referring to a 3-particle
coupling in the general formula of Ref.~\cite{Luscher:1985dn} is
absent. At leading order in the chiral expansion the scattering
amplitude is given by the constant expression
$\Fpipi=-\mpi^2/\fpi^2$. Inserting this into
Eq.~\eqref{eqn:luescher_pion} yields
\begin{eqnarray}
\left.\frac{\mpi(L)-\mpi}{\mpi}\right|_\mathrm{LO} &=&
\frac{3}{8\pi^2} \frac{\mpi^2}{\fpi^2} \frac{K_1(\mpi L)}{\mpi L}
\label{eqn:pion_mass_shift} \\ & \simeq & \frac{3}{4(2\pi)^{3/2}}
\frac{\mpi^2}{\fpi^2} \frac{e^{-\mpi L}}{(\mpi L)^{3/2}},
\label{eqn:pion_mass_shift_asymp}
\end{eqnarray}
where $K_1$ is a modified Bessel function, and the second expression follows
from its asymptotic behavior, $K_1(x)\simeq e^{-x}/\sqrt{x}$, for large
$x$. In addition one can take existing NLO and NNLO chiral corrections to the
infinite-volume amplitude $\Fpipi$ into account and solve
Eq.~\eqref{eqn:luescher_pion} numerically \cite{Colangelo:2003hf}. We will
consider the practical effects of such corrections more closely in
Section~\ref{sec:vol_dep}.

Ref.~\cite{Luscher:1983rk} also quotes a finite-size mass shift
formula for the \emph{nucleon} that can be evaluated if the $\pi N$
scattering amplitude as known from experiment is inserted. In a sense
this formula has been superseded, however, by a recent result derived
from Baryon ChPT in finite volume by the QCDSF-UKQCD collaboration
\cite{AliKhan:2003cu}. Using the infrared regularization scheme
\cite{Becher:1999he} they obtain
\begin{eqnarray}
\label{eqn:delta_Op3}
\Delta_a(L) &=& \frac{3 g_A^2 m_0 \mpi^2}{16\pi^2 \fpi^2} \nn \\
&& \times \int_0^\infty\!dx\;\sideset{}{'}\sum_\vec{n}
K_0\Bigl(\!L\abs{\vec{n}}\sqrt{m_0^2x^2+\mpi^2(1-x)}\,\Bigr) \nn \\
\end{eqnarray}
for the nucleon finite-size mass shift $\Delta m_N(L)$ at $O(p^3)$ in the
$p$-expansion of the chiral Lagrangian. The constants $g_A$ and $\fpi$ are to
be taken in the chiral limit, $m_0$ is the nucleon mass in the chiral limit
and the pion mass $\mpi$ parameterizes the quark mass via the
Gell-Mann-Oakes-Renner relation. The pion decay constant $\fpi$ is normalized
such that its physical value is 92.4~MeV. $K_0$ is a modified Bessel
function, and the sum extends over all spatial 3-vectors $\vec{n}$ with
integer components $n_i$, $i=1,2,3$, except $\vec{n}=\vec{0}$. $n_i$ can be
interpreted as the number of times the pion moves around the lattice in the
$i$-th direction. At $O(p^4)$ an additional contribution to the mass shift
$\Delta m_N(L)$ is given by
\begin{eqnarray}
\label{eqn:delta_Op4}
\Delta_b(L) &=& \frac{3\mpi^4}{4\pi^2\fpi^2} \sideset{}{'}\sum_\vec{n}
\biggl[ (2c_1-c_3)\frac{K_1(\abs{\vec{n}}\mpi L)}{\abs{\vec{n}}\mpi L}
  \nn \\
&& \qquad\qquad\qquad\quad\;\;\: +\,c_2 \frac{K_2(\abs{\vec{n}}\mpi L)}{(\abs{\vec{n}}\mpi L)^2}
  \biggr],
\end{eqnarray}
where $c_1$, $c_2$ and $c_3$ are effective couplings and $K_1$ and
$K_2$ are again modified Bessel functions. The complete QCDSF-UKQCD
result for the nucleon finite-size mass shift at NNLO reads
\begin{equation}
\label{eqn:delta_tot}
m_N(L)-m_N = \Delta_a(L) + \Delta_b(L) + O(p^5).
\end{equation}

To apply this formula to simulated lattice data, in
Ref.~\cite{AliKhan:2003cu} the parameters of the chiral
expansion in \eqref{eqn:delta_Op3} and \eqref{eqn:delta_Op4} are taken
partly from phenomenology and partly from a fit of numerical data for
$m_N$ from relatively fine and large lattices to the (infinite-volume)
$O(p^4)$ formula~\cite{Procura:2003ig}
\begin{eqnarray}
\label{eqn:mNmpisq}
m_N &=& m_0 - 4c_1\mpi^2 - \frac{3g_A^2}{32\pi\fpi^2}\,\mpi^3 \nn \\
    && + \biggl[ e_1^r(\lambda) - \frac{3}{64\pi^2\fpi^2}
  \left(\frac{g_A^2}{m_0}-\frac{c_2}{2}\right) \nn \\
    && - \frac{3}{32\pi^2\fpi^2}\left(\frac{g_A^2}{m_0}-8c_1+c_2+4c_3\right)
  \ln\frac{\mpi}{\lambda} \biggr] \mpi^4 \nn \\
    && + \frac{3g_A^2}{256\pi\fpi^2m_0^2}\,\mpi^5 + O(\mpi^6)
\end{eqnarray}
where the counterterm $e_1^r(\lambda)$ is taken at the renormalization scale
$\lambda$. With all parameters fixed in this way, the formulae
\eqref{eqn:delta_Op3} and \eqref{eqn:delta_Op4} provide parameter-free
predictions of the finite-volume effects in the nucleon
mass. Eq.~\eqref{eqn:delta_tot} has already been demonstrated to work
remarkably well for various volumes at pion masses of around 550 and 700~MeV
\cite{AliKhan:2003cu}, and we will show in Section~\ref{sec:vol_dep} that it
is also capable of describing the volume dependence of our nucleon masses at
pion masses from about 640 down to approximately 420~MeV. The QCDSF-UKQCD
collaboration has shown that if the leading ($\abs{\vec{n}}=1$) terms of their
$O(p^4)$ formula \eqref{eqn:delta_tot} are expressed in a form that
corresponds to L\"uscher's approach \cite{Luscher:1983rk}, his nucleon formula
is essentially recovered.  A remaining numerical discrepancy has recently been
identified as being due to a missing factor of two in the so-called \emph{pole
term} of L\"uscher's nucleon formula \cite{Koma:2004wz}. An important
advantage of the formula \eqref{eqn:delta_tot} is that it is valid not just
asymptotically but also at smaller values of $L$, because its subleading terms
($\abs{\vec{n}}>1$) account also for those virtual pions that cross the
boundary of the lattice more than just once.

Besides the formulae \eqref{eqn:luescher_pion} and
\eqref{eqn:delta_tot} we will confront our data also with the
observation by Fukugita \etal\ that their pion, rho and nucleon masses
from simulations with dynamical staggered quarks followed a power law,
\begin{equation}
\label{eqn:power_law}
\Delta m_H(L) \propto L^{-n} \quad \text{with} \quad n\simeq 2..3,
\end{equation}
rather than L\"uscher's formula
\cite{Fukugita:1992jj,Fukugita:1992hr}. Their result has been
interpreted such that at smaller, sub-asymptotic volumes the leading
finite-size effect originates from a distortion of the hadronic
wave-function itself, contrary to the large-volume picture of a
squeezed cloud of virtual pions surrounding a point-like hadron.

\section{\label{sec:sim_details}Simulation}

A numerical investigation of finite-size effects in LQCD requires gluon field
ensembles from several different lattice volumes at fixed gauge coupling and
quark mass. In order to take benefit from our previous SESAM and \txl\
projects \cite{Eicker:1998sy,Lippert:1997vg} we have conducted supplementary
simulations using again the standard Wilson action, the gauge part of which is
given by the plaquette action
\begin{equation}
\label{eqn:s_g}
S_\mathrm{g} = \beta \sum_x \sum_{\mu<\nu} \left[ 1-\frac{1}{3} \re \tr
  \plaq(x) \right],
\end{equation}
and the quark part by
\begin{equation}
\label{eqn:wilson_quark_action}
S_\mathrm{q} = \sum_{x,y} \qbar(x)M(x,y)q(y),
\end{equation}
where
\begin{eqnarray}
\label{eqn:wilson_dirac_matrix}
M(x,y) &=& \delta_{xy} - \kappa \sum_\mu \Big[ (1-\gamma_\mu) U_\mu(x)
  \delta_{x+\muhat,y} \nn \\ && + (1+\gamma_\mu)
  U_\mu^\dagger(x-\muhat) \delta_{x-\muhat,y} \Big].
\end{eqnarray}
We worked at two different values of the gauge coupling parameter,
$\beta=5.32144$ (with $\kappa=0.1665$) and $\beta=5.6$ (with
$\kappa=0.1575,0.158$). The larger $\beta$ corresponds to the value
used previously by SESAM/\txl. The smaller $\beta$ and the
corresponding $\kappa$ result from linear ex\-tra\-po\-lations of
lines of constant $\mps/\mv$ and $\mps L$ in the
$(\beta,\kappa)$-plane, based on SESAM/\txl\ data and aiming at
$\mps/\mv\lesssim 0.5$ and $\mps L\approx 5$ on a $16^3$-lattice
\cite{Orth:2001mk}. For each of these $\beta$,$\kappa$-combinations we
have produced unquenched gauge field configurations for at least three
different lattice volumes $L^3$ with $L$ varying between 10 and 16,
thus complementing ensembles from SESAM and \txl\ with $L=16$ and
$24$, respectively. Generating the configurations periodic boundary
conditions were imposed in all four space-time directions for the
gauge field, while for the pseudofermions we used periodic boundary
conditions in the spatial directions and antiperiodic boundary
conditions in the temporal direction. Beside the original SESAM TAO
code that was used on a 512-processor APE100/Quadrics (QH4) we worked
with an adapted version of the code on APEmille. There, a
128-processor \textit{crate} was used to generate the $16^3\times
32$-lattices, while the $12^3\times 32$-lattices were produced on a
\textit{unit} of 32 processors. On ALiCE \cite{Eicker:1999gz}, the
128-node ``Alpha Linux Cluster Engine'' at the University of
Wuppertal, an optimized C/MPI-version (with core routines written in
Assembler) \cite{Sroczynski:2002kn,Sroczynski:2003zp} of the SESAM
code ran on partitions of 16 ($12^3\times 32$ and $14^3\times 32$
lattices) and 8 processors ($10^3\times 32$ lattice). All the codes
are implementations of the $\Phi$-version \cite{Gottlieb:1987mq} of
the Hybrid Monte Carlo algorithm for two degenerate quark flavors.

\begin{table*}
 \caption{\label{tab:sim_details_1}The parameters of our simulations
  within the GRAL project together with those of previous SESAM and
  \txl\ runs which we have compiled here for reference (see also
  Table~\ref{tab:sim_details_2}). The chronological start-vector guess
  (csg) has been implemented only in our APE code. Note that in the
  case of ALiCE runs the numbers for $\langle\niter\rangle$
  (\textsl{slanted}) refer to the solve of $M^\dagger Y = \phi$ only,
  whereas in the case of APE runs they refer to the full two-step
  solution of $(M^\dagger M)X=\phi$.  Note also that in order to boost
  the initially low acceptance rate of only 41\% in the simulation at
  $(\beta,\kappa,L) = (5.32144,0.1665,16)$ we decreased both the step
  size $\dt$ and the number of molecular dynamics steps $\nmd$ at some
  intermediate stage of the simulation.}
 \begin{ruledtabular}
  \begin{tabular}{lllcD{.}{.}{1}rldlrl}
   \cc{$\beta$} & \cc{$\kappa$} & \cc{$L^3 T$} & \cc{Precnd.} & \cc{$N_\mathrm{csg}$} &\cc{$\nmd$} &
   \cc{$\dt$} & \cc{$T$} & \cc{$P_\mathrm{acc}$} & \cc{$\langle\niter\rangle$} & \cc{$\langle\,\Box\,\rangle$} \\
   \hline
   5.32144 & 0.1665 & $12^3 32$ & ll-SSOR  & \mbox{-} & $125\pm 20$ & 0.004 & 0.5  & 0.71 & \textsl{147(6)} & 0.53949(14)  \\
           &        & $14^3 32$ & ll-SSOR  & \mbox{-} & $125\pm 20$ & 0.004 & 0.5  & 0.64 & \textsl{130(6)} & 0.53879(15)  \\
           &        & $16^3 32$ & ll-SSOR  &       7  & $200\pm 40$ & 0.005 & 1.0  & 0.41 &                 &              \\
           &        &           &          &          & $125\pm 20$ & 0.004 & 0.5  & 0.65 &         315(9)  & 0.538290(65) \\
   5.5     & 0.1580 & $16^3 32$ & ll-SSOR  &       7  & $100\pm 20$ & 0.010 & 1.0  & 0.77 &          45(1)  & 0.555471(45)  \\
           & 0.1590 & $16^3 32$ & ll-SSOR  &       7  & $100\pm 20$ & 0.010 & 1.0  & 0.71 &          85(1)  & 0.558164(38)  \\
           & 0.1596 & $16^3 32$ & ll-SSOR  &       7  & $100\pm 20$ & 0.010 & 1.0  & 0.61 &         138(2)  & 0.559745(58)  \\
           & 0.1600 & $16^3 32$ & ll-SSOR  &       7  & $100\pm 20$ & 0.010 & 1.0  & 0.40 &         216(3)  & 0.560776(47)  \\
   5.6     & 0.1560 & $16^3 32$ & even/odd &       6  & $100\pm 20$ & 0.010 & 1.0  & 0.82 &          86(1)  & 0.569879(25)  \\
           & 0.1565 & $16^3 32$ & ll-SSOR  &       7  & $100\pm 20$ & 0.010 & 1.0  & 0.77 &          90(1)  & 0.570721(22)  \\
           & 0.1570 & $16^3 32$ & ll-SSOR  &       7  & $100\pm 20$ & 0.010 & 1.0  & 0.67 &         133(1)  & 0.571592(27)  \\
           & 0.1575 & $10^3 32$ & ll-SSOR  & \mbox{-} & $100\pm 20$ & 0.010 & 1.0  & 0.87 &  \textsl{63(1)} & 0.573114(27)  \\
           &        & $12^3 32$ & ll-SSOR  &       7  & $100\pm 20$ & 0.010 & 1.0  & 0.76 &         146(2)  & 0.572771(30)  \\
           &        & $14^3 32$ & ll-SSOR  & \mbox{-} & $100\pm 20$ & 0.010 & 1.0  & 0.62 &  \textsl{79(1)} & 0.572598(22)  \\
           &        & $16^3 32$ & even/odd &      11  & $100\pm 20$ & 0.010 & 1.0  & 0.78 &         293(6)  &               \\
           &        &           & ll-SSOR  &       3  & $ 71\pm 12$ & 0.007 & 0.5  & 0.73 &         160(6)  & 0.572550(27)  \\
           &        & $24^3 40$ & ll-SSOR  &       6  & $125\pm 20$ & 0.004 & 0.5  & 0.80 &         109(1)  & 0.572476(13)  \\
           & 0.1580 & $12^3 32$ & ll-SSOR  &       7  & $125\pm 20$ & 0.008 & 1.0  & 0.85 &         150(5)  & 0.573793(32)  \\
           &        & $14^3 32$ & ll-SSOR  & \mbox{-} & $100\pm 20$ & 0.005 & 0.5  & 0.88 & \textsl{113(1)} & 0.573677(25)  \\
           &        & $16^3 32$ & ll-SSOR  &       7  & $125\pm 20$ & 0.006 & 0.75 & 0.66 &         302(5)  & 0.573461(25)  \\
           &        & $24^3 40$ & ll-SSOR  &       6  & $125\pm 20$ & 0.004 & 0.5  & 0.62 &         256(7)  & 0.573375(16)  \\
  \end{tabular}
 \end{ruledtabular}
\end{table*}

\begin{table*}
 \caption{\label{tab:sim_details_2}Overview of run lengths, thermalization
  times, numbers of equilibrium configurations, numbers of analyzed
  configurations and the distances between them, machines used for the
  generation of the configurations, and the corresponding projects. (See text
  for more details.)}
 \begin{ruledtabular}
  \begin{tabular}{lllrrrlccc}
   \cc{$\beta$} & \cc{$\kappa$} & \cc{$L^3 T$} & \cc{$N_\mathrm{traj}$} &
   \cc{$N_\mathrm{therm}$} & \cc{$N_\mathrm{equi}$} & \cc{$\Delta N_\mathrm{traj}$}
   & \cc{$\nconf$} & \cc{Machine} & \cc{Project} \\
   \hline
   5.32144 & 0.1665 & $12^3 32$ & 10100 & 1500 &  8600 & $50\pm 6$ & 170 & ALiCE        & GRAL  \\
           &        & $14^3 32$ &  5900 &  800 &  5100 & $40\pm 4$ & 129 & ALiCE        & GRAL  \\
           &        & $16^3 32$ & 15300 & 8600 &  6700 & $40\pm 4$ & 169 & APE100/mille & GRAL  \\           
   5.5     & 0.1580 & $16^3 32$ &  4000 & 1000 &  3000 & $25\pm 3$ & 119 & APE100       & SESAM \\
           & 0.1590 & $16^3 32$ &  6000 & 1000 &  5000 & $25\pm 3$ & 200 & APE100       & SESAM \\
           & 0.1596 & $16^3 32$ &  5500 &  500 &  5000 & $25\pm 3$ & 199 & APE100       & SESAM \\
           & 0.1600 & $16^3 32$ &  5500 &  500 &  5000 & $25\pm 3$ & 200 & APE100       & SESAM \\ 
   5.6     & 0.1560 & $16^3 32$ &  5700 &  600 &  5100 & $25$      & 198 & APE100       & SESAM \\
           & 0.1565 & $16^3 32$ &  5900 &  700 &  5200 & $24$      & 208 & APE100       & SESAM \\
           & 0.1570 & $16^3 32$ &  6000 & 1000 &  5000 & $25$      & 201 & APE100       & SESAM \\
           & 0.1575 & $10^3 32$ & 16000 & 2600 & 13400 & $48\pm 4$ & 278 & ALiCE        & GRAL  \\
           &        & $12^3 32$ &  8000 &  700 &  7300 & $30\pm 4$ & 243 & APEmille     & GRAL  \\
           &        & $14^3 32$ &  8400 & 1400 &  7000 & $30\pm 4$ & 231 & ALiCE        & GRAL  \\
           &        & $16^3 32$ &  6500 & 1400 &  5100 & $25$      & 206 & APE100       & SESAM \\
           &        & $24^3 40$ &  5100 &  500 &  4600 & $25$      & 185 & APE100       & \txl  \\
           & 0.1580 & $12^3 32$ &  3000 &  500 &  2500 & $24\pm 2$ & 103 & APEmille     & GRAL  \\
           &        & $14^3 32$ &  9100 & 1300 &  7800 & $40\pm 4$ & 195 & ALiCE        & GRAL  \\
           &        & $16^3 32$ &  6500 & 1100 &  5400 & $30\pm 4$ & 181 & APEmille     & GRAL  \\
           &        & $24^3 40$ &  4500 &  700 &  3800 & $24$      & 158 & APE100       & \txl  \\
  \end{tabular}
 \end{ruledtabular}
\end{table*}

Tables~\ref{tab:sim_details_1} and \ref{tab:sim_details_2} give an overview of
the production runs we conducted within the GRAL project. For reference and
convenience we also list the corresponding figures for previous SESAM/\txl\
runs here and in the following tables. The SESAM/\txl\ simulations at
$(\beta,\kappa,L)=(5.6,0.1575,16)$, $(5.6,0.1575,24)$ and $(5.6,0.158,24)$
have been used for our analysis of finite-size effects. Except for some early
SESAM runs (or parts thereof) featuring an even/odd decomposition of the quark
matrix $M$, in all simulations the BiCGStab algorithm was used with
locally-lexicographic SSOR preconditioning
\cite{Frommer:1994vn,Fischer:1996th} for the inversion of the full matrix. The
linear system $(M^\dagger M)X=\phi$ was solved in two steps: first $M^\dagger
Y = \phi$ was solved for $Y$ and then $MX=Y$ was solved for
$X$. $\langle\niter\rangle$ in Table~\ref{tab:sim_details_1} denotes the
average numbers of iterations the solver needed until convergence. Note that
the \textsl{slanted} numbers quoted for runs on ALiCE refer to the solve of
$M^\dagger Y = \phi$ only, whereas in the case of runs on APE they refer to
the full two-step solution of $(M^\dagger M)X=\phi$. For a comparison a
relative factor of approximately two must therefore be taken into
account. Like in the SESAM/\txl\ simulations the stopping accuracy
$R\equiv\norm{MX-\phi}/\norm{X}$ was chosen to be $R=10^{-8}$ in all GRAL
runs.

Our APE programs additionally feature an implementation of the
chronological start vector guess proposed in
Ref.~\cite{Brower:1995vx}. In our GRAL simulations the depth of
the extrapolation, $N_\mathrm{csg}$, was not optimized, however, but
rather fixed to 7.

Both for decreasing quark mass and increasing lattice volume (all
other parameters constant, respectively) we observe a drop in the
acceptance rate as anticipated.

Table~\ref{tab:sim_details_2} lists the total number of generated
trajectories, $N_\mathrm{traj}$, the first $N_\mathrm{therm}$ of which
we attribute to the thermalization phase and therefore discard, so
that we are left with $N_\mathrm{equi}$ equilibrium configurations,
respectively. (In the thermalization phase of each production run we
approached the respective target quark mass adiabatically from larger
quark masses. These initial trajectories are in general not counted
here. An exception to this rule is the run at $(\beta,\kappa,L) =
(5.32144,0.1665,16)$ where a rather long initial tuning phase
incorporated several changes of the simulation parameters.) $\nconf$
configurations out of these, separated by $\Delta N_\mathrm{traj}$ (up
to a uniform, random variation) trajectories, have been analyzed
further.

\begin{table}
 \caption{\label{tab:autocorr}Measured autocorrelation times for the
  average number of solver iterations, $\niter$, and the plaquette,
  $\Box$, for those (parts of) runs in which the fermion matrix was
  ll-SSOR preconditioned (see Table~\ref{tab:sim_details_1}).}
 \begin{ruledtabular}
  \newcolumntype{(}{D{(}{(}{3}}
  \begin{tabular}{lll((((}
  \cc{$\beta$} & \cc{$\kappa$} & \cc{$L^3 T$}
  & \cc{$\tauexp^{\niter}$} & \cc{$\tauint^{\niter}$}
  & \cc{$\tauexp^{\Box}$} & \cc{$\tauint^{\Box}$} \\
  \hline
  5.32144 & 0.1665 & $12^3 32$ & 52( 8) &  58( 5) & 53( 9) & 50( 3) \\
          &        & $14^3 32$ & 45( 7) &  39( 3) & 46( 7) & 26( 2) \\
          &        & $16^3 32$ & 97(14) & 105( 7) & 77( 9) & 54( 4) \\
          &        &           & 94(22) &  77(12) & 74(10) & 57( 5) \\
  5.5     & 0.1580 & $16^3 32$ & 21( 5) &  20( 1) & 27( 3) & 16( 1) \\
          & 0.1590 & $16^3 32$ & 25( 7) &  25( 1) & 31( 3) & 13( 1) \\
          & 0.1596 & $16^3 32$ & 74(12) &  38( 2) & 37(11) & 17( 1) \\
          & 0.1600 & $16^3 32$ & 56( 6) &  46( 3) & 64( 7) & 34( 2) \\
  5.6     & 0.1560 & $16^3 32$ &        &         &        &        \\
          & 0.1565 & $16^3 32$ & 29( 6) &  19( 4) &  7( 1) &  5( 1) \\
          & 0.1570 & $16^3 32$ & 35( 6) &  25( 5) &  9( 3) &  6( 1) \\
          & 0.1575 & $10^3 32$ & 62( 7) &  28( 2) & 15( 3) &  5( 1) \\
          &        & $12^3 32$ & 24( 3) &  20( 1) & 15( 3) &  6( 1) \\
          &        & $14^3 32$ & 88(27) &  34( 3) & 26( 8) &  8( 1) \\
          &        & $16^3 32$ & 47( 7) &  33( 4) & 18( 6) &  7( 4) \\
          &        & $24^3 40$ & 51( 7) &  36( 4) & 11( 2) &  7( 3) \\
          & 0.1580 & $12^3 32$ & 25( 5) &  24( 1) &  9( 3) &  4( 1) \\
          &        & $14^3 32$ & 14( 2) &  12( 1) & 10( 1) &  4( 1) \\
          &        & $16^3 32$ & 43(15) &  24( 2) & 14( 4) &  8( 2) \\
          &        & $24^3 40$ & 61(19) &  50( 5) & 20(10) & 20( 2) \\
  \end{tabular}
 \end{ruledtabular}
\end{table}

\subsection*{Autocorrelation}

A suitable estimator of the true autocorrelation (or
\emph{autocovariance}) function for a finite time-series $A_t$,
$t=1,\dots,\tmc$, is given by
\begin{equation}
C^A(t) = \frac{1}{\tmc-t} \sum_{s=1}^{\tmc-t} \left( A_s -
  \ev{A}_{\!L} \right) \left( A_{s+t} - \ev{A}_{\!R} \right),
\end{equation}
where the use of the ``left'' and ``right'' mean-value estimators 
\begin{equation}
\ev{A}_{\!L} = \frac{1}{\tmc-t}\!\sum_{r=1}^{\tmc-t}\! A_r,\;\;\;
\ev{A}_{\!R} = \frac{1}{\tmc-t}\!\sum_{r=1}^{\tmc-t}\! A_{r+t}
\end{equation}
in general leads to a faster convergence of $C^A(t)$ to the true
autocorrelation function for \mbox{$\tmc\to\infty$}
\cite{Lippert:habil}. From fits of the estimator of the normalized
autocorrelation function,
\begin{equation}
\rho^A(t) = C^A(t)/C^A(0),
\end{equation}
to an exponential we extract estimates for the exponential
autocorrelation time $\tauexp^A$, defined as
\begin{equation}
\label{eqn:tauexp}
\tauexp^A = \limsup_{t\to\infty}\frac{t}{-\log \abs{\rho^A(t)}},
\end{equation}
for $A=\Box,\niter$. Due to the notorious difficulty of determining
autocorrelation times from short time-series we refrained from an
elaborate optimization of the fit ranges, and the values given in
Table~\ref{tab:autocorr} should be considered as rough estimates of
the exponential autocorrelation times only. We have checked, however,
that the differencing method described in
Ref.~\cite{Lippert:habil} yields consistent results. The
measured values for $\tauexp^A$ are generally larger than the
integrated autocorrelation times $\tauint^A$ that we measure as usual
with the help of Sokal's ``windowing'' procedure and which are also
shown in the table. We use the finite sum
\begin{equation}
\label{eqn:tauint}
\tauint^A = \frac{1}{2} + \sum_{t=1}^{\tcut} \rho^A(t)
\end{equation}
with a variable cut-off $\tcut$ to estimate $\tauint^A$. Plotting the
resulting values against $\tcut$ does, ideally, reveal a plateau for
$\tcut\to\tmc$. If a plateau does not emerge, we typically either find
a maximum, or $\tauint^A(\tcut)$ is monotonously rising. If there is a
maximum, we choose the corresponding value as best estimate of
$\tauint^A$. Otherwise we reverse Sokal's proposal to choose $\tcut$
larger than 4 to 6 times $\tauint^A$: we assume $\tauint^A$ to lie in
the interval defined by the intersections of the straight lines with
slopes $\tcut/4$ and $\tcut/6$, respectively, with the curve
$\tauint^A(\tcut)$.

Comparing autocorrelation times for runs with different lattice
volumes we find only a weak increase of the autocorrelation times with
increasing volume. More striking is the difference in the
autocorrelation times between the simulations at $\beta=5.6$ and
$\beta=5.32144$. At the stronger coupling the relatively large
autocorrelation times reflect the long-ranged statistical fluctuations
that we observe in the corresponding time series. These fluctuations
are more severe on the smaller lattices where, moreover, zero modes of
the Dirac matrix start playing a role. On the largest volume at this
$\beta$ the situation is somewhat better: While the autocorrelation
times are comparable to those on the smaller volumes, we see no
indication of exceptional configurations on this ($16^3$) lattice.

\section{\label{sec:observables}Physical observables}

\subsection{\label{sec:potential}Static quark potential}

We calculated the static quark potential in order to determine the
Sommer parameter \cite{Sommer:1993ce} that we use to set the physical
scale. For the SESAM/\txl\ runs at $\beta=5.5$ and $\beta=5.6$ the
Sommer radii $R_0\equiv r_0/a$ as listed in Table~\ref{tab:r0} have
been published previously in Refs.~\cite{Eicker:2001dn} and
\cite{Bali:2000vr}, respectively. Since the lattice-size
dependence of $R_0$ is assumed to be small and as we want to have a
common length scale for the different simulated lattice volumes at a
given gauge coupling and quark mass, we have adopted the $R_0$-value
from the largest available lattice, respectively, also for the smaller
ones.

In order to determine the Sommer radius for the $16^3$ lattice at
$(\beta,\kappa)=(5.32144,0.1665)$ we measured the Wilson loops $W(R,\tau)$
with temporal extents of up to $\tau=8$ and spatial separations of up to
$R=\sqrt{3}\cdot 7\approx 12$ lattice units on the same configurations that we
used for spectroscopy. We employed the modified Bresenham algorithm of
Ref.~\cite{Bolder:2000un} to include all possible lattice vectors
$\vec{R}$ for a given separation $R\equiv\abs{\vec{R}}$. Using the spatial APE
smearing as described in Ref.~\cite{Albanese:1987ds}, we applied
\begin{equation}
\mathrm{link} \to \alpha \times \mathrm{link} +  \mathrm{staples}
\end{equation}
to the gauge links of each configuration before calculating the Wilson
loops. We used $\alpha=2.3$ and performed $N=26$ iterations, followed by
a projection back into the gauge group~\cite{Bali:1992ab}.

The asymptotic behavior of the static potential $V(R)$ for
sufficiently large times $\tau$ is given by
\begin{equation}
W(R,\tau) \sim C(R)\,e^{-V(R)\tau},
\end{equation}
so that one can define the effective potential
\begin{equation}
V_\mathrm{eff}(R,\tau) = \ln \frac{W(R,\tau)}{W(R,\tau+1)}.
\end{equation}

\begin{figure}
\includegraphics*[width=\columnwidth]{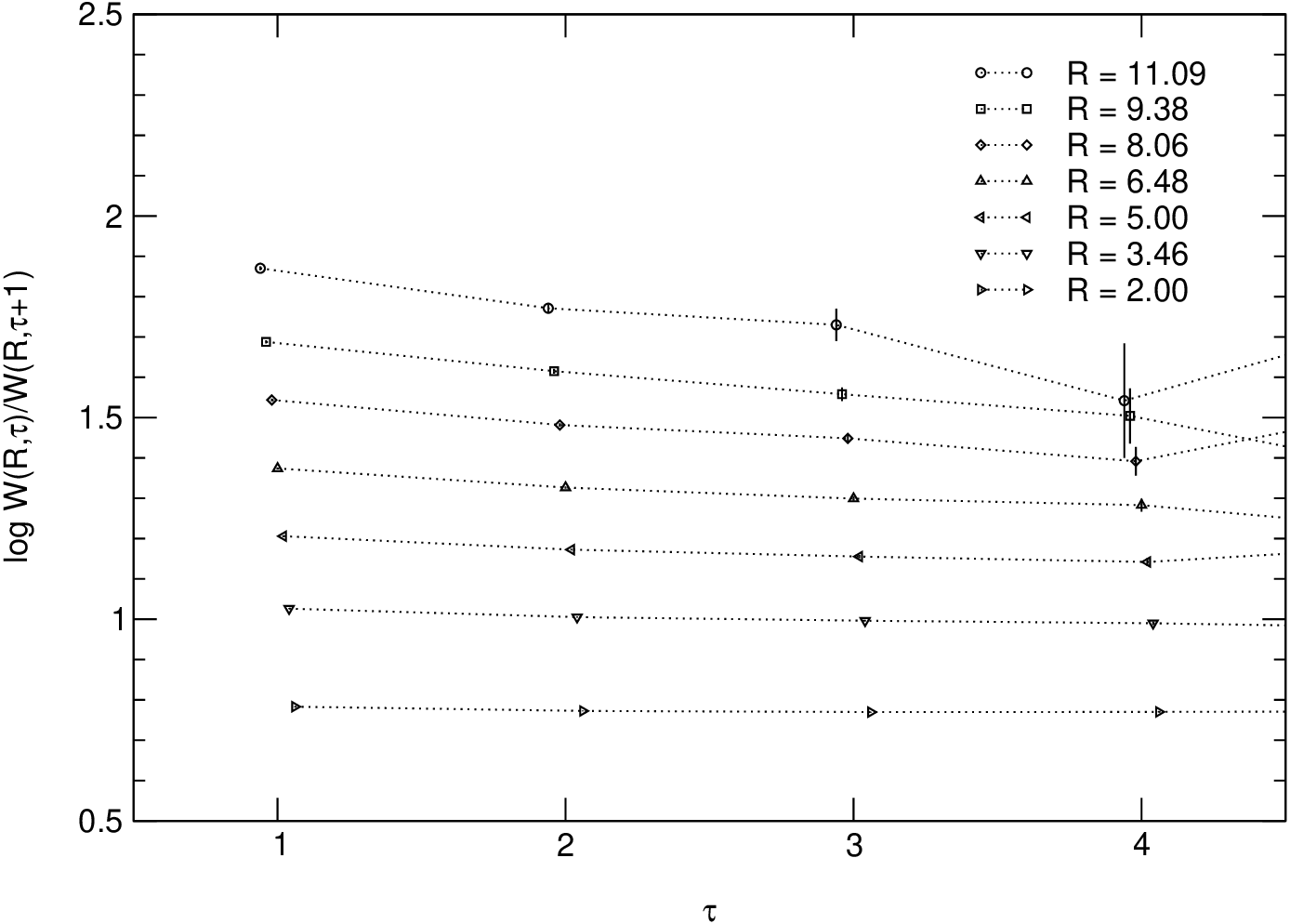}
\caption{\label{fig:Veff}The effective potential,
  $V_\mathrm{eff}(R,\tau)$, at $(\beta,\kappa)=(5.32144,0.1665)$ on
  the $16^3$ lattice, for selected values of $R$ in the range
  $\tau=1,\dots,4$. Larger values of $\tau$ are dominated by
  sta\-ti\-sti\-cal noise.}
\end{figure}

\begin{figure}
\includegraphics*[width=\columnwidth]{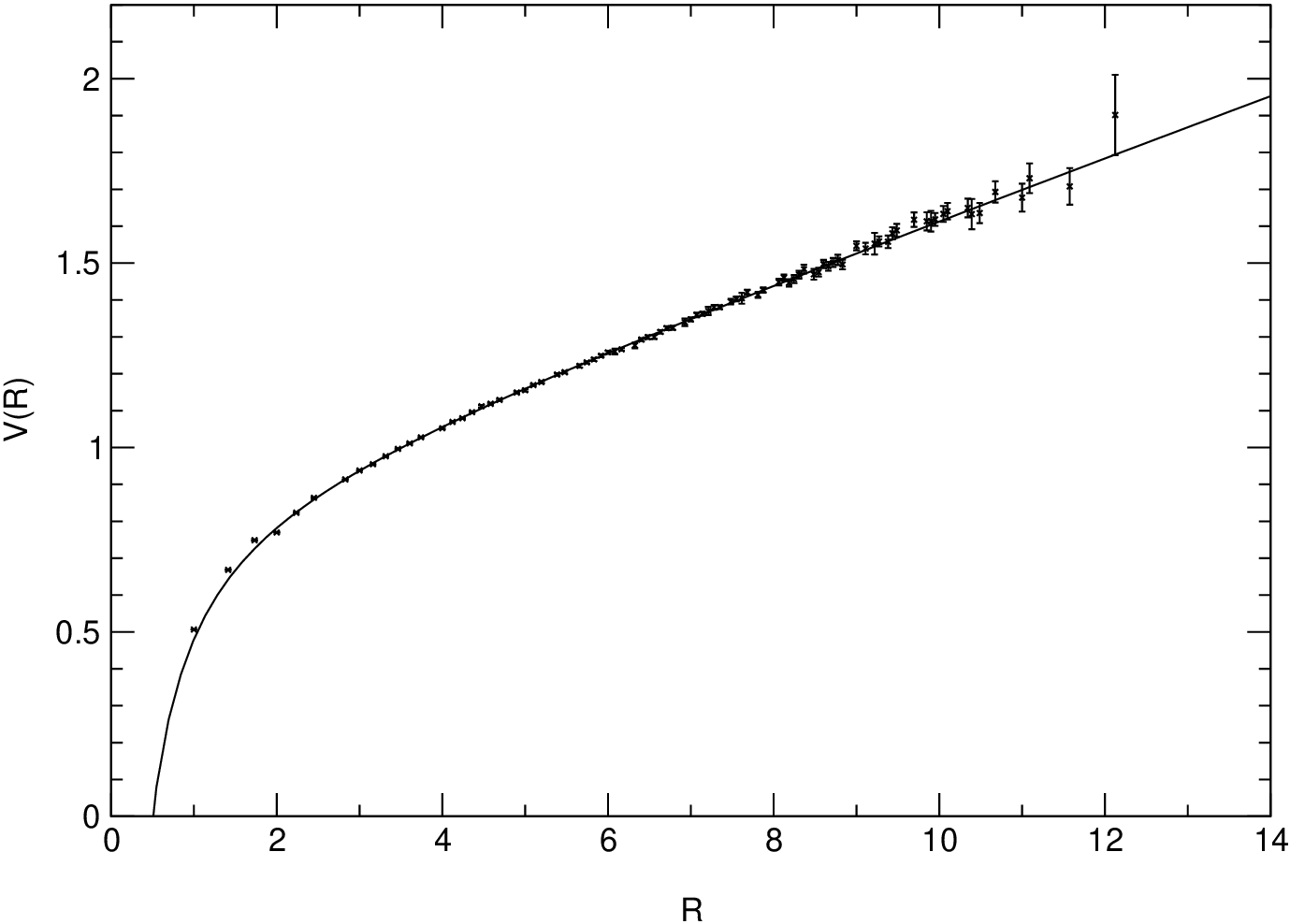}
\caption{\label{fig:pot}The static quark potential obtained from
  Wilson loops at $(\beta,\kappa)=(5.32144,0.1665)$ on the $16^3$
  lattice.}
\end{figure}

Fig.~\ref{fig:Veff} shows the $\tau$-dependence of the effective
potential $V_\mathrm{eff}(R,\tau)$ for various values of $R$. At
$\tau=3$ the effective potential is already largely independent of
$\tau$ while the statistical errors are moderate, so that we determined
$V(R)$ from a single exponential fit in the range
$[\tau_\mathrm{min},\tau_\mathrm{max}]=[3,4]$.
As can be seen from Fig.~\ref{fig:pot}, the resulting values for
$V(R)$ show the expected behavior.
We observe no indication of string breaking \cite{Bali:2004pb} and therefore
fit the data in the range $[R_\mathrm{min},R_\mathrm{max}]=[2.5,8]$ to
\begin{equation}
V(R) = V_0 - \frac{e}{R} + \sigma R.
\end{equation}
The upper boundary of the fit range was set to $R_\mathrm{max}=8$ because up
to this value the data correspond nicely to the expected linear behavior, with
small statistical errors. With $R_\mathrm{max}$ held fixed the lower boundary
was determined by investigating the $R_\mathrm{min}$-dependence of $r_0$ for
various values $R_\mathrm{min}\gtrsim 2$. (Below this value one observes a
violation of rotational symmetry due to the finite lattice spacing.)  From the
fit we obtained the following parameters for the potential (in lattice units):
\[
V_0 = 0.8378(87), \quad e = 0.440(18), \quad \sigma = 0.08187(97).
\]

The Sommer scale $R_0$, which is defined through the force between two
static quarks at some intermediate distance,
\begin{equation}
\label{eqn:sommer_radius}
R^2 \left.\frac{dV}{dR}\right|_{R=R_0} = 1.65,
\end{equation}
was obtained from these parameters according to
\begin{equation}
R_0 = \sqrt{\frac{1.65-e}{\sigma}}.
\end{equation}

Our result for the simulation at
$(\beta,\kappa,L)=(5.32144,0.1665,16)$ is given in
Table~\ref{tab:r0}. The quoted uncertainty corresponds to the
statistical error.

\begin{table*}
 \caption{\label{tab:r0}Sommer scale and resulting momentum cut-off,
  lattice spacing and lattice size for $r_0 = 0.5$~fm.}
 \begin{ruledtabular}
  \begin{tabular}{lllllll}
   \cc{$\beta$} & \cc{$\kappa$} & \cc{$L^3T$} & \cc{$r_0/a$} &
   \cc{$a^{-1}\,[\mathrm{GeV}]$} & \cc{$a\,[\mathrm{fm}]$} & \cc{$L\,[\mathrm{fm}]$} \\
   \hline
   5.32144 & 0.1665 & $16^3 32$ & 3.845(37) & 1.517(15)  & 0.1300(13)  & 2.081(21)  \\
   5.5     & 0.1580 & $16^3 32$ & 4.027(24) & 1.5893(95) & 0.12416(74) & 1.987(12)  \\
           & 0.1590 & $16^3 32$ & 4.386(26) & 1.731(11)  & 0.11400(68) & 1.824(11)  \\
           & 0.1596 & $16^3 32$ & 4.675(34) & 1.845(14)  & 0.10695(78) & 1.711(13)  \\
           & 0.1600 & $16^3 32$ & 4.889(30) & 1.929(12)  & 0.10227(63) & 1.636(10)  \\
   5.6     & 0.1560 & $16^3 32$ & 5.104(29) & 2.014(12)  & 0.09796(56) & 1.5674(89) \\
           & 0.1565 & $16^3 32$ & 5.283(52) & 2.085(21)  & 0.09464(93) & 1.514(15)  \\
           & 0.1570 & $16^3 32$ & 5.475(72) & 2.161(29)  & 0.0913(12)  & 1.461(20)  \\
           & 0.1575 & $16^3 32$ & 5.959(77) & 2.352(31)  & 0.0839(11)  & 1.343(18)  \\
           &        & $24^3 40$ & 5.892(27) & 2.325(11)  & 0.08486(39) & 2.0367(94) \\
           & 0.1580 & $24^3 40$ & 6.230(60) & 2.459(24)  & 0.08026(78) & 1.926(19)  \\
  \end{tabular}
 \end{ruledtabular}
\end{table*}

We used $r_0 = 0.5$~fm to set the scale in our simulations. The resulting
physical values for the momentum cut-off $a^{-1}$, the lattice spacing $a$ and
the box size $L$ are displayed in Table~\ref{tab:r0}. The smallest simulated
$\beta$ of 5.32144 is associated with a lattice spacing of $a=0.13$~fm,
corresponding to a momentum cut-off of 1.52~GeV. While we have to watch out
for potentially large $O(a)$ discretization errors at this coupling, the
physical volume is the biggest of all simulated volumes. With a linear
extension of slightly more than 2~fm it is comparable in size with the \txl\
lattice at $(\beta,\kappa)=(5.6,0.1575)$.

\subsection{\label{sec:masses}Hadron masses and decay amplitudes}

\begin{figure}
\includegraphics*[width=\columnwidth]{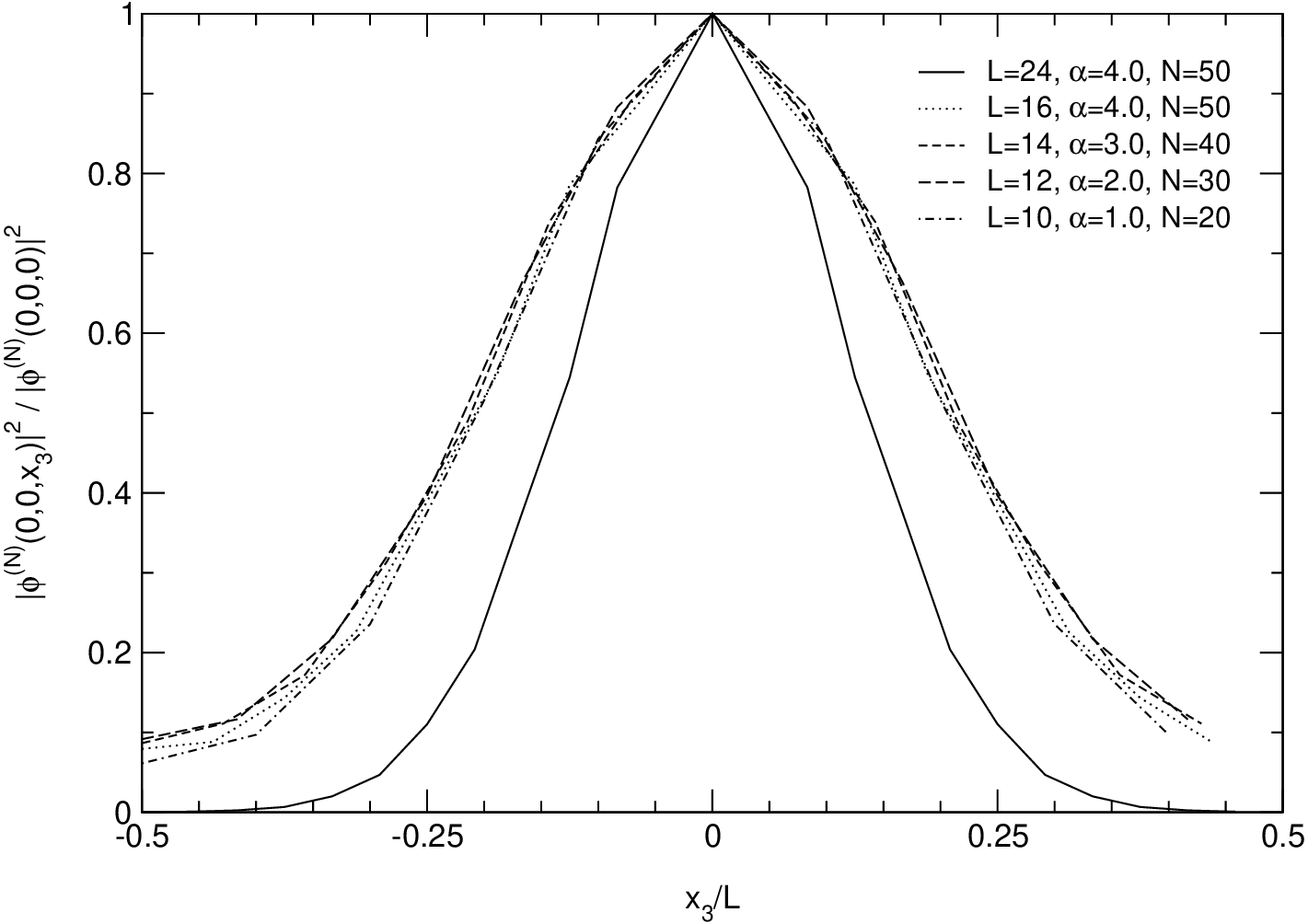}
\caption{\label{fig:smearing}The parameters $N$, $\alpha$ for the Wuppertal
  smearing scheme were chosen such as to yield approximately the same wave
  function shapes for both the smaller lattices and the SESAM lattice
  ($L=16$).}
\end{figure}

In order to extract meson masses and amplitudes we followed the standard
procedure of computing zero-momentum 2-point functions ($x=(\xvec,\tau)$)
\begin{equation}
\label{eqn:two_point_fcn}
\ev{\Ocal^\dagger(\tau)\Ocal(0)} \equiv \sum_{\xvec} \ev{\Ocal^\dagger(x)\Ocal(0)}
\end{equation}
for the following pseudoscalar and vector operators:
\begin{subequations}\label{eqn:operators}
\begin{align}
P(x) & =\qbar^\prime(x)\gamma_5 q(x), \label{eqn:op_ps}\\
A_4(x) & =\qbar^\prime(x)\gamma_5\gamma_4 q(x), \label{eqn:op_ax}\\
V_k(x) & =\qbar^\prime(x)\gamma_k q(x). \label{eqn:op_v}
\end{align}
\end{subequations}
For the nucleon we have used the octet baryon operator
\begin{equation}
N(x)=\epsilon_{abc}\bigl(q^T_a(x) C \gamma_5
q^\prime_b(x)\bigr)q_c^{\prime\prime}(x),
\end{equation}
where $a,b,c$ are color indices and $C=\gamma_4\gamma_2$ is the charge
conjugation matrix. We employed the gauge invariant Wuppertal smearing
\cite{Gusken:1989qx} at the source only ($ls$) or at both source and
sink ($ss$).  In the previous SESAM and \txl\ simulations $N=50$
smearing steps were used with a weight of $\alpha=4.0$. These
parameters were originally optimized for the $16^3$ SESAM lattice and
then adopted for the larger $24^3$ \txl\ lattice, too. In order to
adapt these parameters to our smaller lattices we have investigated
the effect of smearing on the various volumes. We applied the
Wuppertal smearing procedure to point sources $\phi^{(0)}(\xvec)$ of
size $L^3$ with $L=24,16,14,12,10$. We set $\phi^{(0)}(\xvec)\equiv 0$
except for the point at $(L/2,L/2,L/2)$, which we defined as the
origin of the respective lattice and where we set
$\phi^{(0)}(\vec{0})=1$. Applying the smearing prescription $N$ times
to $\phi^{(0)}$ with all $U_\mu(\xvec)\equiv 1$ we plot the amplitude
of the resulting wave function $\phi^{(N)}$ along the (arbitrarily
chosen) $(0,0,1)$-direction relative to its maximum at the origin,
\ie\ $\abs{\phi^{(N)}(0,0,x_3)}^2/\abs{\phi^{(N)}(\vec{0})}^2$, versus
$x_3/L$, for various values of $N$ and $\alpha$. On inspection of the
resulting wave function shapes we selected the parameters $N$ and
$\alpha$ for our simulated volumes so as to make the respective wave
function profile look approximately like the SESAM one. The chosen
smearing parameters are listed in Table~\ref{tab:smearing_params},
while the corresponding wave function profiles are displayed in
Fig.~\ref{fig:smearing}.

\begin{table}
 \caption{\label{tab:smearing_params}Smearing parameters.}
 \begin{ruledtabular}
  \begin{tabular}{cccccc}
  $L$      & 10  & 12  & 14  & 16  & 24  \\ \hline
  $\alpha$ & 1.0 & 2.0 & 3.0 & 4.0 & 4.0 \\
  $N$      & 20  & 30  & 40  & 50  & 50  \\
  \end{tabular}
 \end{ruledtabular}
\end{table}

The masses and amplitudes of mesons were obtained by correlated
least-$\chi^2$ fits of the (time-symmetrized) correlators
$\ev{\Ocal^\dagger(\tau) \Ocal(0)}$, $\Ocal=P,A_4,V_k$, to the
parameterization
\begin{equation}
\label{eqn:meson_fit_fcn}
f^M_\tau(C,M) = C \left( e^{-M\tau} + e^{-M(T-\tau)} \right),
\end{equation}
where $M\equiv am$ is the mass and
$C\sim|\braopket{0}{\Ocal}{p}|^2/2M$ if $\ket{p}$ is the zero-momentum
state of the particle associated with $\Ocal$. In the case of the
nucleon we fitted the correlator (anti-symmetrized in $\tau$ and
$T-\tau$) to the single exponential
\begin{equation}
\label{eqn:baryon_fit_fcn}
f^B_\tau(C,M) = C e^{-M\tau}.
\end{equation}
The optimal lower limits of the fit intervals, $\taumin$, were found
as usual by examining the $\chisqdof$-behavior and the stability of
the masses with respect to $\taumin$, and by inspection of the
effective masses. The upper limit, $\taumax$, was in general kept
fixed at $T/2$ for mesons and $T/2-1$ for the nucleon.

\begin{table*}
 \caption{\label{tab:latt_masses}Masses of the pseudoscalar and vector
  mesons and of the nucleon, the pseudoscalar decay constant and the
  PCAC quark mass in lattice units.}
 \begin{ruledtabular}
  \begin{tabular}{lllllllll}
   \cc{$\beta$} & \cc{$\kappa$} & \cc{$L^3 T$} & \cc{$\mps L$} &
   \cc{$\Mps$} & \cc{$\Mv$} & \cc{$M_N$} & \cc{$Z_A^{-1}\Fps$} & \cc{$Z_q M_q$} \\
   \hline
   5.32144 & 0.1665 & $12^3 32$ & 3.18(8) & 0.2648(67) & 0.508(26)  & 0.788(26)  & 0.062(10)  & 0.0106(33)  \\
           &        & $14^3 32$ & 3.6(1)  & 0.2577(87) & 0.518(12)  & 0.779(16)  & 0.0757(56) & 0.0152(18)  \\
           &        & $16^3 32$ & 4.42(7) & 0.2760(42) & 0.4999(78) & 0.727(11)  & 0.0843(62) & 0.0155(12)  \\
   5.5     & 0.1580 & $16^3 32$ & 8.85(6) & 0.5534(39) & 0.6506(46) & 1.026(18)  & 0.1073(42) & 0.0821(35)  \\
           & 0.1590 & $16^3 32$ & 7.09(4) & 0.4429(26) & 0.5529(54) & 0.8718(78) & 0.0945(23) & 0.0544(12)  \\
           & 0.1596 & $16^3 32$ & 5.89(4) & 0.3682(27) & 0.4902(52) & 0.7640(75) & 0.0815(16) & 0.03724(81) \\
           & 0.1600 & $16^3 32$ & 4.89(5) & 0.3058(34) & 0.4547(61) & 0.703(10)  & 0.0750(23) & 0.0279(15)  \\
   5.6     & 0.1560 & $16^3 32$ & 7.15(4) & 0.4469(23) & 0.5365(36) & 0.8533(62) & 0.0843(19) & 0.0620(11)  \\
           & 0.1565 & $16^3 32$ & 6.32(6) & 0.3948(38) & 0.4989(54) & 0.785(10)  & 0.0805(18) & 0.0467(15)  \\
           & 0.1570 & $16^3 32$ & 5.52(5) & 0.3452(29) & 0.4527(52) & 0.7095(90) & 0.0726(16) & 0.0391(15)  \\
           & 0.1575 & $10^3 32$ & 4.92(6) & 0.4919(55) & 0.587(20)  & 1.042(20)  & 0.0284(30) & 0.0209(32)  \\
           &        & $12^3 32$ & 4.3(1)  & 0.3576(89) & 0.494(12)  & 0.817(16)  & 0.0429(34) & 0.0249(19)  \\
           &        & $14^3 32$ & 4.27(6) & 0.3048(44) & 0.4413(66) & 0.719(16)  & 0.0566(30) & 0.0261(22)  \\
           &        & $16^3 32$ & 4.49(6) & 0.2806(35) & 0.4036(68) & 0.6254(89) & 0.0626(26) & 0.0275(16)  \\
           &        & $24^3 40$ & 6.64(6) & 0.2765(26) & 0.3944(38) & 0.5920(75) & 0.0646(18) & 0.02680(68) \\
           & 0.1580 & $12^3 32$ & 4.6(1)  & 0.387(12)  & 0.535(17)  & 0.882(25)  & 0.022(12)  & 0.0113(88)  \\
           &        & $14^3 32$ & 4.13(8) & 0.2949(60) & 0.4677(90) & 0.717(19)  & 0.0233(32) & 0.0099(17)  \\
           &        & $16^3 32$ & 3.72(8) & 0.2325(51) & 0.371(13)  & 0.622(12)  & 0.0469(26) & 0.0141(11)  \\
           &        & $24^3 40$ & 4.78(8) & 0.1991(33) & 0.3519(86) & 0.500(12)  & 0.0602(39) & 0.0157(11)  \\
  \end{tabular}
 \end{ruledtabular}
\end{table*}

\begin{table*}
 \caption{\label{tab:phys_masses}Masses of the pseudoscalar meson, the
  vector meson and the nucleon in physical units (using
  $r_0=0.5$~fm).}
 \begin{ruledtabular}
  \begin{tabular}{lllllllll}
   \cc{$\beta$} & \cc{$\kappa$} & \cc{$L^3 T$} & \cc{$L\,[\mathrm{fm}]$} &
   \cc{$(r_0\mps)^2$} & \cc{$\mps/\mv$} & \cc{$\mps\,[\mathrm{GeV}]$} & \cc{$\mv\,[\mathrm{GeV}]$} & \cc{$m_N\,[\mathrm{GeV}]$} \\
   \hline
   5.32144 & 0.1665 & $12^3 32$ & 1.56(2)  & 1.037(57) & 0.521(23)  & 0.402(11)  & 0.771(40)  & 1.195(42) \\
           &        & $14^3 32$ & 1.82(2)  & 0.982(68) & 0.497(20)  & 0.391(14)  & 0.786(20)  & 1.182(26) \\
           &        & $16^3 32$ & 2.08(2)  & 1.126(43) & 0.552(11)  & 0.4188(75) & 0.759(14)  & 1.104(20) \\
   5.5     & 0.1580 & $16^3 32$ & 1.99(1)  & 4.97(20)  & 0.8506(31) & 0.8795(81) & 1.0340(95) & 1.631(30) \\
           & 0.1590 & $16^3 32$ & 1.82(1)  & 3.77(12)  & 0.8010(53) & 0.7666(64) & 0.957(11)  & 1.509(16) \\
           & 0.1596 & $16^3 32$ & 1.71(1)  & 2.96(10)  & 0.7512(51) & 0.6793(70) & 0.904(12)  & 1.410(17) \\
           & 0.1600 & $16^3 32$ & 1.64(1)  & 2.235(85) & 0.6725(93) & 0.5901(75) & 0.877(13)  & 1.356(21) \\
   5.6     & 0.1560 & $16^3 32$ & 1.567(9) & 5.20(18)  & 0.8330(16) & 0.9002(69) & 1.0807(94) & 1.719(16) \\
           & 0.1565 & $16^3 32$ & 1.51(1)  & 4.35(25)  & 0.7912(72) & 0.823(11)  & 1.040(15)  & 1.637(27) \\
           & 0.1570 & $16^3 32$ & 1.46(2)  & 3.57(21)  & 0.7627(58) & 0.746(12)  & 0.978(17)  & 1.533(28) \\
           & 0.1575 & $10^3 32$ & 0.849(4) & 8.40(59)  & 0.838(30)  & 1.144(14)  & 1.365(48)  & 2.424(47) \\
           &        & $12^3 32$ & 1.018(5) & 4.44(48)  & 0.724(11)  & 0.832(21)  & 1.149(27)  & 1.901(38) \\
           &        & $14^3 32$ & 1.188(5) & 3.22(18)  & 0.691(11)  & 0.709(11)  & 1.026(16)  & 1.671(39) \\
           &        & $16^3 32$ & 1.358(6) & 2.73(12)  & 0.6952(99) & 0.6524(86) & 0.938(16)  & 1.454(22) \\
           &        & $24^3 40$ & 2.037(9) & 2.654(90) & 0.7010(62) & 0.6429(67) & 0.9171(98) & 1.377(19) \\
           & 0.1580 & $12^3 32$ & 0.963(9) & 5.80(88)  & 0.722(20)  & 0.951(30)  & 1.316(44)  & 2.167(66) \\
           &        & $14^3 32$ & 1.12(1)  & 3.37(28)  & 0.630(13)  & 0.725(16)  & 1.150(25)  & 1.763(50) \\
           &        & $16^3 32$ & 1.28(1)  & 2.10(15)  & 0.627(21)  & 0.572(14)  & 0.912(33)  & 1.530(32) \\
           &        & $24^3 40$ & 1.93(2)  & 1.539(74) & 0.566(17)  & 0.4896(94) & 0.865(23)  & 1.228(31) \\
  \end{tabular}
 \end{ruledtabular}
\end{table*}

The masses (in lattice units) of the pseudoscalar and vector mesons
and of the nucleon are listed in Table~\ref{tab:latt_masses}. As the
masses from the $ls$ and the $ss$ correlators are consistent we only
quote the values extracted from the latter. The quoted errors are
statistical in nature and have been estimated with the jackknife
method (after suitable blocking of the
data). Table~\ref{tab:latt_masses} also shows $\mps(L) L$, the linear
box size in units of the pseudoscalar correlation length $1/\mps(L)$,
where $\mps(L)$ is the pion mass in the given \emph{finite} volume. It
should be borne in mind that for sub-asymptotic volumes this value is
in general significantly different from $\mps L$, where $\mps$ is the
pseudoscalar mass in \emph{infinite} volume. At
$(\beta,\kappa)=(5.32144,0.1665)$ we attain our lightest quark mass,
with $\mps/\mv$ being close to 0.5. Using $r_0=0.5$~fm to set the
physical scale the hadron masses of Table~\ref{tab:latt_masses}
translate into the values listed in Table~\ref{tab:phys_masses}.
This table shows also the physical box sizes $L\,[\mathrm{fm}]$ which have
been calculated using the lattice spacing $a$ from the largest available
lattice, respectively (see also Table~\ref{tab:r0}). The dimensionless
quantity
\begin{equation}
M_r = (r_0 \mps)^2
\end{equation}
is another measure of the quark mass, since for $m_q\to 0$ the pion
mass behaves like $\mpi \propto \sqrt{m_q}$. At the physical strange
quark mass it gives $M_r\approx 3.1$ \cite{Farchioni:2002vn}. At
those parameter sets where we have simulated several lattice volumes
the value of $M_r$ ranges between $M_r\approx 2.65$ and $M_r\approx
1.13$, corresponding to about 85 and 36\% of the value for the strange
quark mass.

The (unrenormalized) pseudoscalar decay constant, which is defined on
the lattice by (for $\pvec=\vec{0}$)
\begin{equation}
Z_A \, \braopket{0}{A_4}{\mathrm{PS}} = \Mps\Fps,
\end{equation}
has been obtained from
\begin{equation}
\label{eqn:Fps}
Z_A^{-1} \Fps = \sqrt{ \frac{2C_A^{ll}}{\Mps} } 
= C_A^{ls} \sqrt{ \frac{2}{\Mps C_A^{ss}} },
\end{equation}
where we have used the fact that amplitudes for local source and sink ($ll$)
can be obtained from $ls$ and $ss$ amplitudes according to
\begin{equation}
\label{eqn:local_amplitudes}
C^{ll} = \frac{(C^{ls})^2}{C^{ss}}.
\end{equation}
The pseudoscalar mass $\Mps=M_P^{ss}$ in Eq.~\eqref{eqn:Fps} has been
taken from a fit of $\ev{P^\dagger(\tau)P(0)}^{ss}$, the amplitudes
$C_A^{ls}$ ($C_A^{ss}$) from a fit of the local-smeared
(smeared-smeared) correlator $\ev{A_4^\dagger(\tau)A_4(0)}^{ls(ss)}$.

In order to determine the (unrenormalized) quark mass $M_q\equiv a
m_q$ as defined via the PCAC relation on the lattice,
\begin{equation}
M_q = -\frac{\Mps}{2} \frac{Z_A}{Z_P}
\frac{\braopket{0}{A_4}{\mathrm{PS}}}{\braopket{0}{P}{\mathrm{PS}}},
\end{equation}
we used the relation
\begin{equation}
Z_q M_q = \frac{\Mps}{2} \sqrt{ \frac{C_A^{ll}}{C_P^{ll}} } =
\frac{\Mps}{2} \frac{C_A^{ls}}{C_P^{ls}} \sqrt{\frac{C_P^{ss}}{C_A^{ss}}},
\end{equation}
where the renormalization constant is defined as $Z_q \equiv Z_P/Z_A$
and the pseudoscalar mass is taken to be the average
\begin{equation}
\Mps = \frac{1}{4} \left( M_P^{ls}+M_P^{ss}+M_A^{ls}+M_A^{ss} \right).
\end{equation}

Our results (in lattice units) for the unrenormalized pseudoscalar
decay constant $\Fps/Z_A$ are displayed in
Table~\ref{tab:latt_masses}. The normalization of the pseudoscalar
decay constant is such that the physical value is $\fpi=92.4$~MeV. The
same table also shows the results for the bare PCAC quark mass
$Z_qM_q$.

\section{\label{sec:vol_dep}Volume dependence of pion and nucleon
  masses}

The three parameter sets at which we have data from several lattice volumes,
namely $(\beta,\kappa)=(5.6,0.1575)$, $(5.6,0.158)$ and $(5.32144,0.1665)$,
are characterized by the quark mass, which in turn can be expressed in terms
of the pion mass via the GMOR relation. We quote the pion mass measured on the
largest lattice, respectively, when we refer to a particular simulation point
$(\beta,\kappa)$. We have investigated the volume dependence of the pion, the
rho and the nucleon at pion masses (before continuum extrapolation) of
approximately 643~MeV, 490~MeV and 419~MeV in the ranges 0.85--2.04~fm,
0.96--1.93~fm and 1.56--2.08~fm, respectively. Due to angular momentum
conservation the decay $\rho\to\pi\pi$ is suppressed on small lattices where
the minimum non-zero momentum $2\pi/L$ is large. We therefore incorporate the
rho resonance in our phenomenological analysis of finite-size effects, because
on the lattices considered here it should be stable.

\begin{figure}
\includegraphics*[width=\columnwidth]{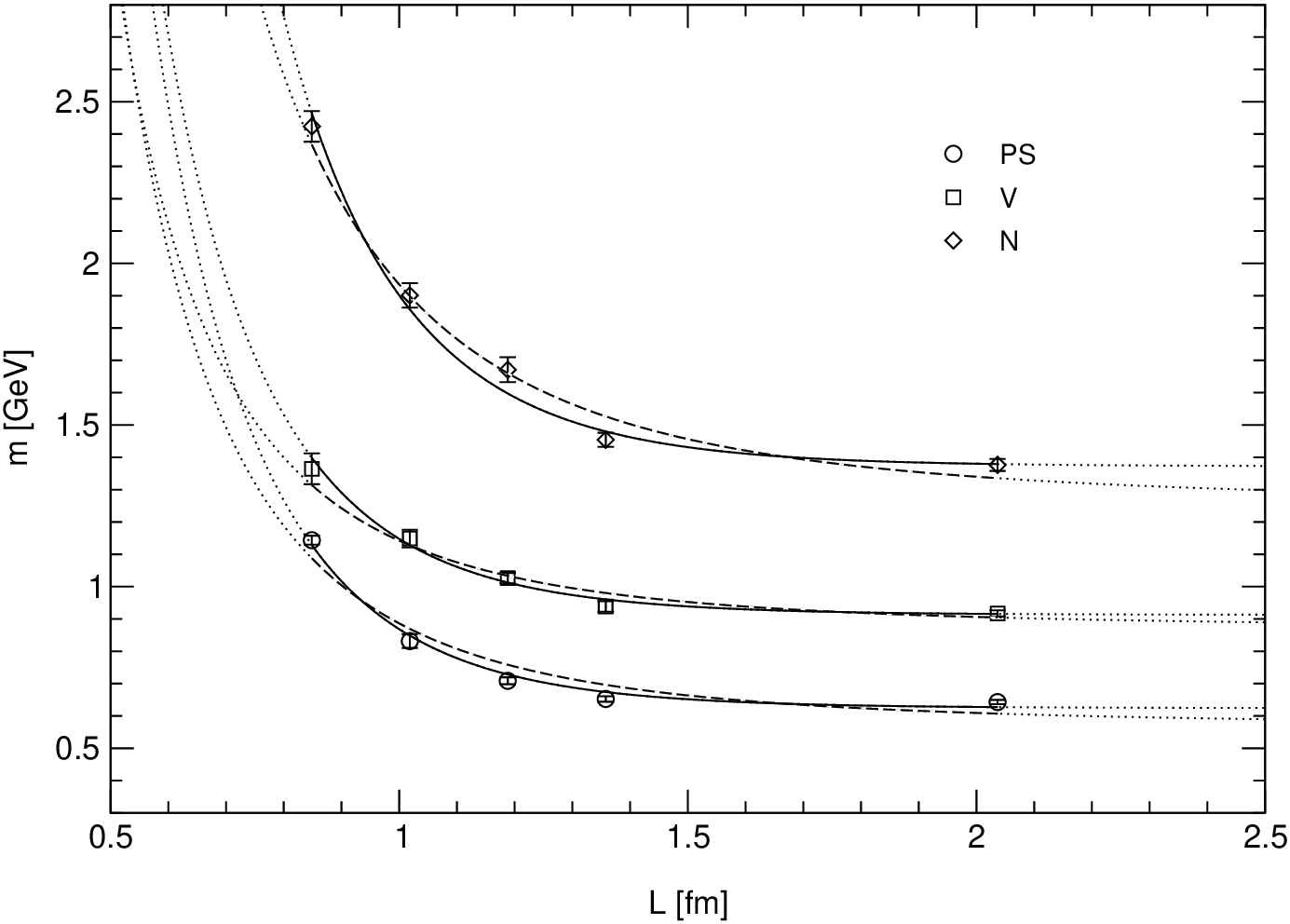}
\caption{\label{fig:ldep_56_1575}Box-size dependence of the pseudoscalar and
  vector meson masses and of the nucleon mass at
  $(\beta,\kappa)=(5.6,0.1575)$. The solid lines result from fits to an
  exponential, Eq.~\eqref{eqn:exp}, while the dashed lines represent fits to a
  power law, Eq.~\eqref{eqn:pow}. The curves are dotted outside the fit
  interval.}
\end{figure}

\begin{figure}
\includegraphics*[width=\columnwidth]{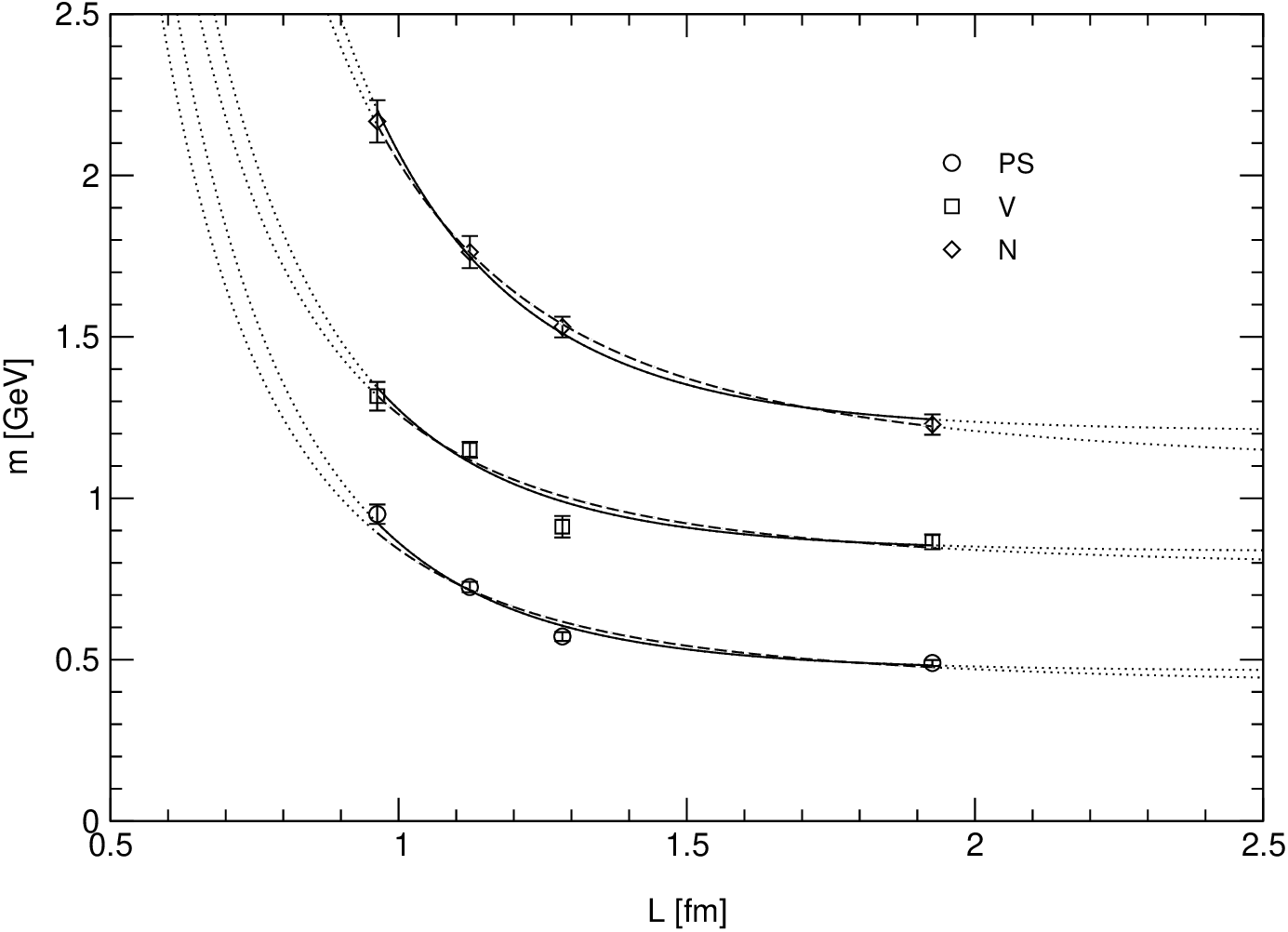}
\caption{\label{fig:ldep_56_1580}Fits as in Fig.~\ref{fig:ldep_56_1575}
  for $(\beta,\kappa)=(5.6,0.158)$.}
\end{figure}

\begin{figure}
\includegraphics*[width=\columnwidth]{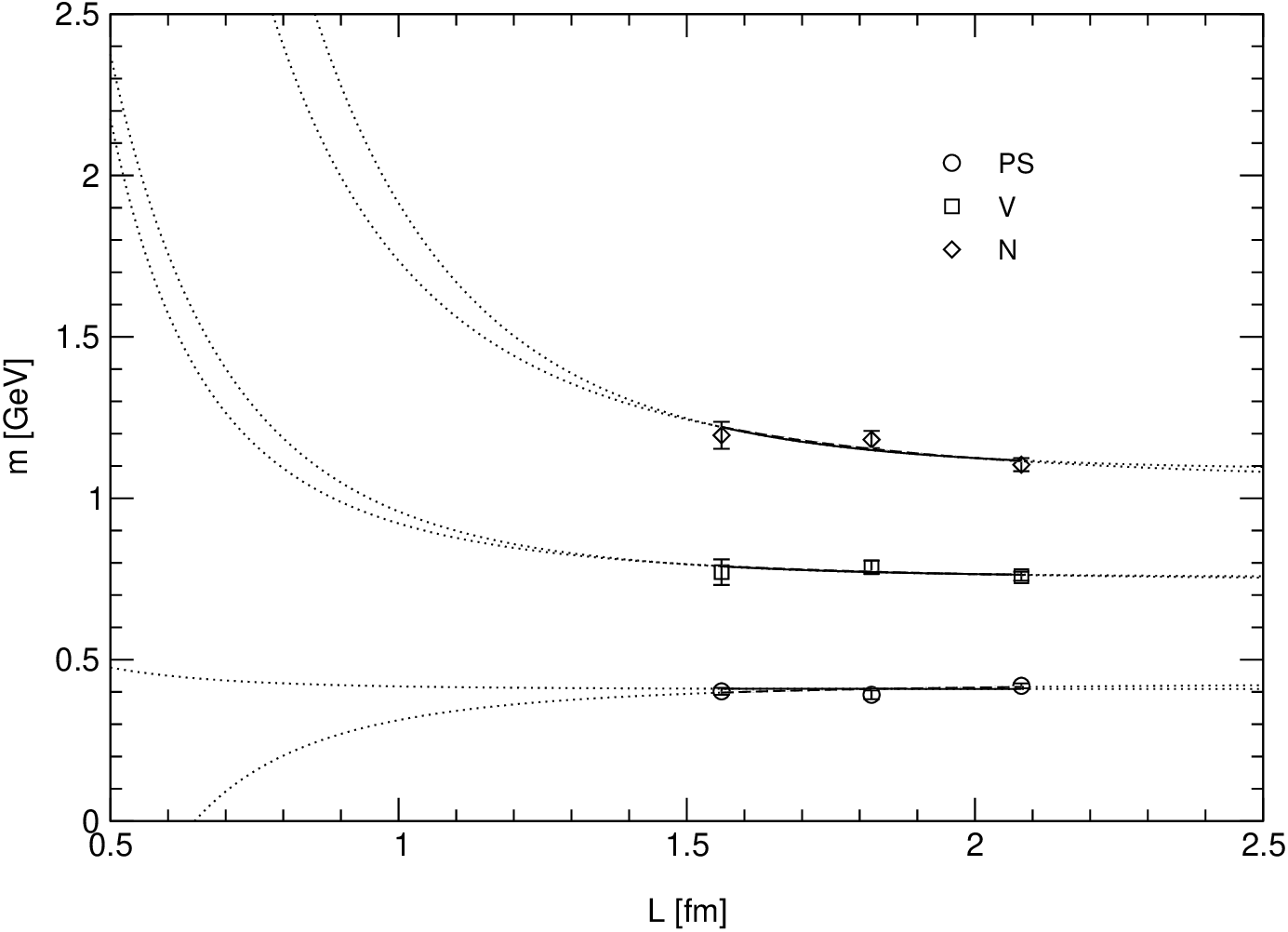}
\caption{\label{fig:ldep_532144_1665}Fits as in
  Fig.~\ref{fig:ldep_56_1575} for
  $(\beta,\kappa)=(5.32144,0.1665)$. The fact that the pion and the
  rho show no monotonous increase of the finite-size shift towards
  decreasing box-size is ascribed to the smallness of the effect and
  statistics. (The simulations of the smaller lattices in particular
  at $\beta=5.32144$ were affected by sizeable fluctuations.) In the
  case of the nucleon the finite-size effect is somewhat more
  significant.}
\end{figure}

\begin{table}
 \caption{\label{tab:delta}Ratios of the pseudoscalar mass and the
  chiral symmetry breaking scale, and the relative finite-size effects
  according to Eq.~\eqref{eqn:rel_mass_diff}.}
 \begin{ruledtabular}
  \begin{tabular}{lllD{.}{.}{7}lll}
   \cc{$\beta$} & \cc{$\kappa$} & \cc{$L^3 T$} & 
   \cc{$\D\frac{\Mps}{4\pi Z_A^{-1}\Fps}$} &
   \cc{$\Rps$} & \cc{$\Rv$} & \cc{$R_N$} \\
   \hline
   5.32144 & 0.1665 & $12^3 32$ & 0.34(6)   & -0.04(3) & 0.02(5) & 0.08(4) \\
           &        & $14^3 32$ & 0.27(2)   & -0.07(3) & 0.04(3) & 0.07(3) \\
           &        & $16^3 32$ & 0.26(2)   & \phantom{-}0  & 0  & 0  \\
   5.6     & 0.1575 & $10^3 32$ & 1.4(1)    & \phantom{-}0.78(3) & 0.49(5) & 0.76(4) \\
           &        & $12^3 32$ & 0.66(6)   & \phantom{-}0.29(3) & 0.25(3) & 0.38(3) \\
           &        & $14^3 32$ & 0.43(2)   & \phantom{-}0.10(2) & 0.12(2) & 0.21(3) \\
           &        & $16^3 32$ & 0.36(2)   & \phantom{-}0.01(2) & 0.02(2) & 0.06(2) \\
           &        & $24^3 40$ & 0.341(10) & \phantom{-}0       & 0       & 0       \\
           & 0.1580 & $12^3 32$ & 1.4(8)    & \phantom{-}0.94(7) & 0.52(6) & 0.76(7) \\
           &        & $14^3 32$ & 1.0(1)    & \phantom{-}0.48(4) & 0.33(4) & 0.44(5) \\
           &        & $16^3 32$ & 0.39(2)   & \phantom{-}0.17(3) & 0.05(5) & 0.25(4) \\
           &        & $24^3 40$ & 0.26(2)   & \phantom{-}0       & 0       & 0       \\
  \end{tabular}
 \end{ruledtabular}
\end{table}

Figs.~\ref{fig:ldep_56_1575}--\ref{fig:ldep_532144_1665} show, for our
three different quark masses, the pion, rho and nucleon masses in
physical units as functions of the box-size. In Table~\ref{tab:delta}
we list the relative differences of the masses measured at $L$ and
$L_\mathrm{max}$,
\begin{equation}
\label{eqn:rel_mass_diff}
R_H(L) = \frac{M_H(L)-M_H(L_\mathrm{max})}{M_H(L_\mathrm{max})},
\end{equation}
where $H=\mathrm{PS}, \mathrm{V}, \mathrm{N}$ and $L_\mathrm{max}=24$
($L_\mathrm{max}=16$) for $\beta=5.6$ ($\beta=5.32144$). For both quark masses
at $\beta=5.6$ we find a large variation of the hadron masses over the
considered range of lattice sizes. While the finite-size effects in the pion,
rho and nucleon masses are relatively small (of the order of a few percent) if
one compares only the two largest lattices at $\kappa=0.1575$, they rapidly
grow on the smaller volumes ($\sim 50$--$80\%$ at $L=10$). The rate of the
increase is hadron dependent: while at large $L$ the pion has the smallest
relative finite-size effect, the relative shift in the pion mass grows
strongest with decreasing $L$, until it exceeds the effect in the rho mass
from $L=12$ and that in the nucleon mass from $L=10$ downwards. Considering
the finite-size effects at $\kappa=0.158$ (corresponding to a lower quark
mass) we notice that at a given value of $L$ the finite-size effects are
generally much larger at $\kappa=0.158$ than at $0.1575$. Again we observe
that the pion is subject to the strongest relative effect in the regime of
small volumes. Finally, at $(\beta,\kappa)=(5.32144,0.1665)$ (corresponding to
the lightest of our quark masses), we find rather small finite-size effects of
only a few percent in the simulated $L$-range, for all considered hadrons. In
view of the small $\mps L$ values (see Table~\ref{tab:phys_masses}) this is
quite remarkable: if the finite-size effects would only depend on $\mps L$ we
would expect the effects at $\beta=5.32144$ to be of about the same order of
magnitude as those at the smaller volumes at $\beta=5.6$. On the other hand,
due to the large lattice spacing the volumes at $\beta=5.32144$ are, in terms
of the physical size, comparable with the larger volumes at $\beta=5.6$. This
strongly suggests that there are, in fact, different mechanisms responsible
for the observed mass shifts, and that the transition between them is in our
case characterized by the absolute physical lattice size rather than the
product $\mps L$. Quite independently of the pion mass the region in $L$ where
finite-size shifts start to become large is located at around 1.5~fm in our
simulations.

\subsection{\label{sec:fits}Fits of the volume dependence}

First we attempted to describe the volume dependence of our simulated
masses phenomenologically by fitting them to two different
parameterizations. One of these parameterizations is inspired by
L\"uscher's exponential leading-order mass shift formula
\eqref{eqn:pion_mass_shift_asymp} for the pion, while the other one is
directly given by the power law observed by Fukugita \etal,
Eq.~\eqref{eqn:power_law}. Although neither of these approaches can
\textit{a priori} be expected to be valid over the entire range of
considered lattice volumes, and although L\"uscher's formula strictly
speaking has no free fit parameters, on practical grounds it is still
interesting whether based on either of the two functional forms an
empirical description can be found that connects small and
medium-sized volumes to the asymptotic regime.

\begin{figure}
\includegraphics*[width=\columnwidth]{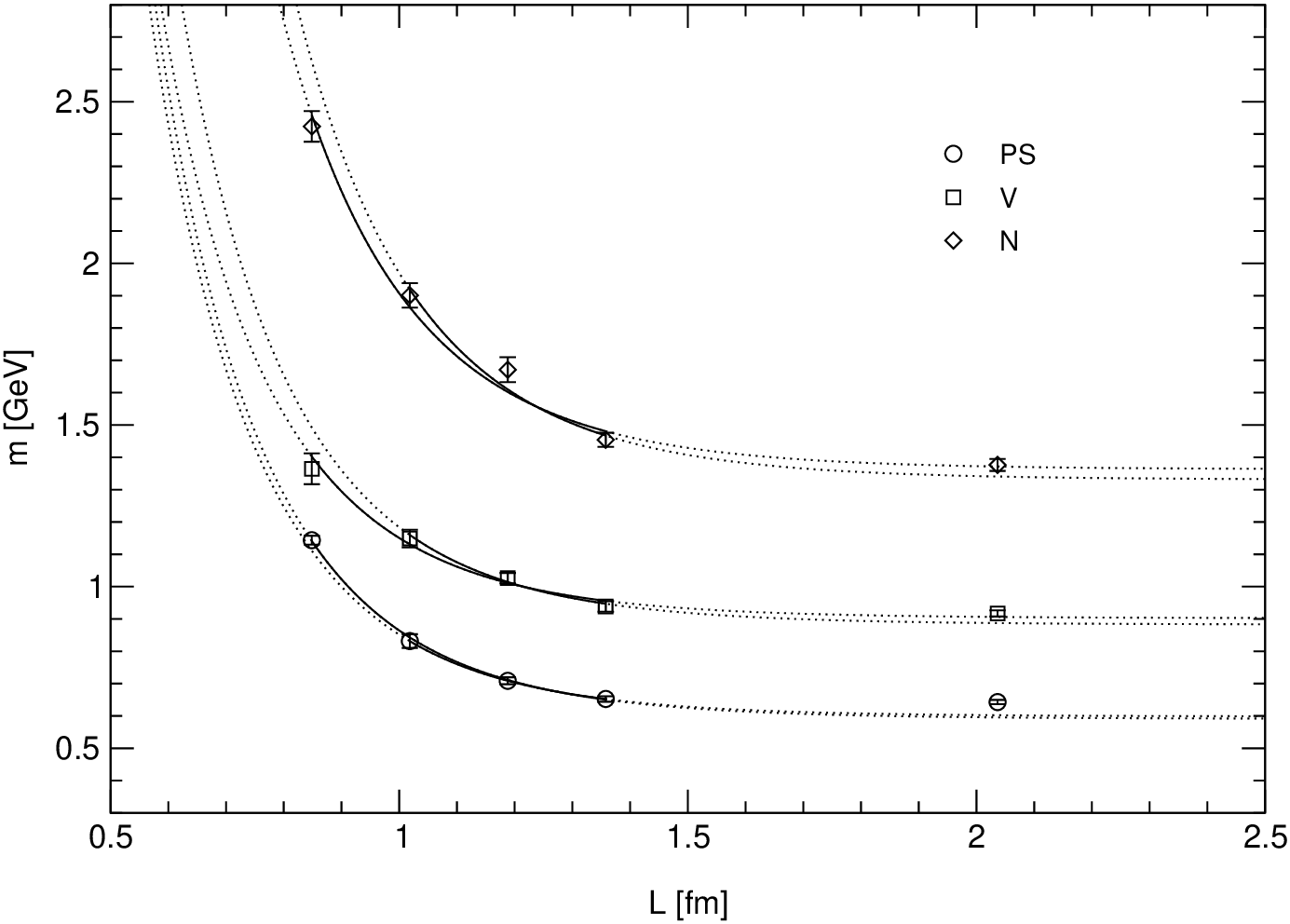}
\caption{\label{fig:exp_par_extra_varl1_56_1575} Infinite-volume extrapolation
  of the masses at $(\beta,\kappa)=(5.6,0.1575)$. The solid lines correspond
  to exponential fits according to Eq.~\eqref{eqn:exp}. The curves are dotted
  outside the fit intervals.}
\end{figure}

\begin{figure}
\includegraphics*[width=\columnwidth]{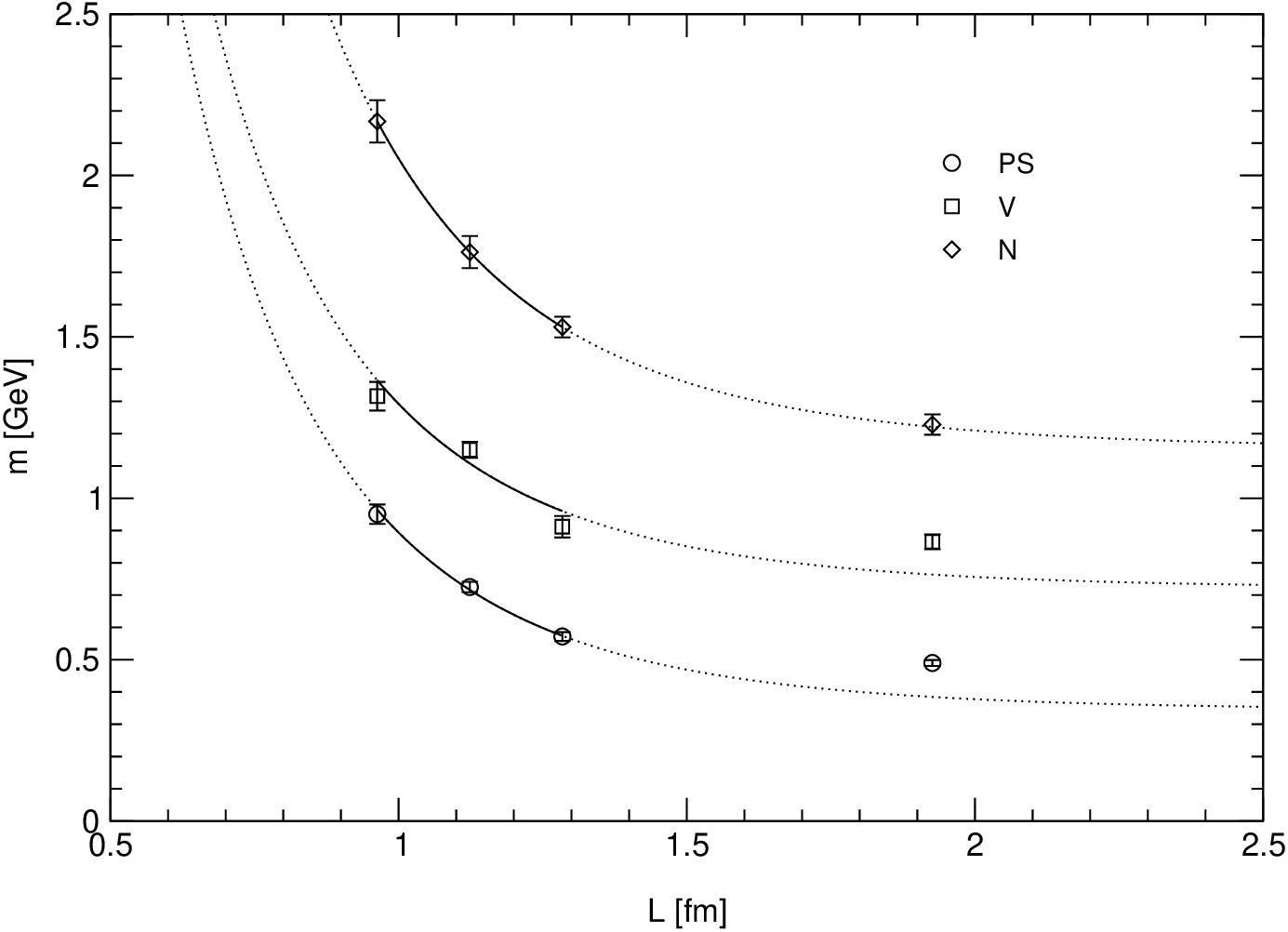}
\caption{\label{fig:exp_par_extra_varl1_56_1580} Extrapolations as in
  Fig.~\ref{fig:exp_par_extra_varl1_56_1575} for
  $(\beta,\kappa)=(5.6,0.158)$.}
\end{figure}

The curves in Figs.~\ref{fig:ldep_56_1575} and \ref{fig:ldep_56_1580} show
fits of the data for the pion, rho and nucleon masses ($H=\text{PS,V,N}$) to
the exponential function
\begin{equation}
\label{eqn:exp}
m_H(L) = m_H + c L^{-\frac{3}{2}} e^{-\mps L},
\end{equation}
and, for comparison, to the power law
\begin{equation}
\label{eqn:pow}
m_H(L) = m_H + c L^{-3}.
\end{equation}
In the case of the pion (where $m_H=\mps$) the mass $\mps$ in
Eq.~\eqref{eqn:exp} was treated as a fit parameter; the result was used as
input for the fits of the rho and the nucleon data, so that all the fits
displayed in Figs.~\ref{fig:ldep_56_1575} and \ref{fig:ldep_56_1580} had two
free parameters. As can be seen from the plots both parameterizations describe
the data reasonably well within the fit interval, but regarding the asymptotic
behavior the exponential ansatz is clearly superior. At
$(\beta,\kappa)=(5.6,0.1575)$, for example, all infinite-volume masses $m_H$
resulting from the exponential fits are compatible with the data from the
largest, $24^3$ lattice, which are assumed to bear no significant finite-size
effects. In contrast, fitting the data to the power law yields numbers for
$m_H$ that grossly underestimate the true asymptotic masses. Varying the right
boundary of the fit interval we find that for small box-sizes ($L \lesssim
1.5$~fm) where the finite-size effects are of the order of several percent the
power law provides an acceptable description of the data. At the two larger
quark masses this corresponds to the regime of $\mps L \lesssim 4.5 .. 4.8$,
in accordance with the common rule of thumb that only for $\mps L\gtrsim 5$
finite-size mass shifts are exponentially suppressed. (In the light of our
results at $(\beta,\kappa)=(5.32144,0.1665)$ it appears as if this rule of
thumb could be relaxed as long as as the physical lattice extent $L$ remains
sufficiently large.) We find that as soon as we include data from larger
volumes (where the mass shifts are small) into the fits the exponential ansatz
yields better values for both $\chisqdof$ and $m_H$. In order to test whether
this ansatz is suitable for an extrapolation from the small lattices to the
infinite volume we fitted the masses at $(\beta,\kappa)=(5.6,0.1575)$ and
$(5.6,0.158)$ only up to $L=16$.  Assuming that the finite-size effects on the
$24^3$ lattice are not significant it can be seen from
Figs.~\ref{fig:exp_par_extra_varl1_56_1575} and
\ref{fig:exp_par_extra_varl1_56_1580} that the asymptotic pion masses are
generally underestimated considerably, while in the case of the nucleon mass
the extrapolation works rather well. In either case the the extrapolation
tends to yield lower bounds to the infinite-volume masses, the systematic
uncertainty of which can be estimated by varying the boundaries of the fit
intervals.

It should be mentioned that alternative fit formulae (obtained \eg\ by
changing the exponent of $L$ from $-3/2$ to $-1$ in
Eq.~\eqref{eqn:exp}, introducing an additional variable factor in the
exponential of Eq.~\eqref{eqn:exp}, or treating the exponent of $L$ as
a free parameter in Eq.~\eqref{eqn:pow}) may also be used to describe
the data. They do not, however, lead to significant improvements
and/or require even more free parameters.

The main lesson we have learned from this exercise is the following: If one
has got hadron masses from more than two different lattice volumes (at fixed
coupling and quark mass) and wants to estimate the infinite-volume masses on
the basis of a fit, one should use an exponential ansatz rather than the power
law. Extrapolating the exponential fit will produce lower bounds to the true
asymptotic masses, and these bounds are generally better than those that can
be obtained with the power law.

\subsection{\label{sec:chpt}Applicability of Chiral Perturbation Theory}

In order to better understand why it is problematic to ex\-tra\-po\-late from
small volumes to the infinite volume on the basis of the simple formulae
\eqref{eqn:exp} and \eqref{eqn:pow} one needs to appreciate their respective
origin and scope. The power law \eqref{eqn:pow} is supposed to originate from
a distortion of the hadron wave function (or from a modification of the effect
of virtual particles traveling around the lattice by a model-dependent form
factor that accounts for the finite hadron extent \cite{Fukugita:1992jj}) at
quite small volumes. Consequently, the $L^{-3}$-behavior is not expected to
persist towards large volumes, which is in fact borne out by our data. On the
other hand, the formula \eqref{eqn:exp} essentially corresponds to L\"uscher's
asymptotic formulae for the pion. L\"uscher's general formula for the volume
dependence of stable particles in a finite volume represents the \emph{leading
term} of a large $L$ expansion, meaning that whenever the relative suppression
factor
\begin{equation}
\frac{\text{subleading}}{\text{leading}} =
O\left(e^{-(\bar{m}-\mpi)L}\right)
\end{equation}
is not small, subleading effects may be of practical relevance.

\subsubsection{\label{sec:pion}Pion}

In the case of the pion we also rely on effective field theory to
provide us with an analytic expression for the elastic forward
scattering amplitude $\Fpipi$. At leading order in the chiral
expansion this amplitude is given by the constant expression
$\Fpipi=-\mpi^2/\fpi^2$. Inserting this into the L\"uscher formula
eventually leads to Eq.~\eqref{eqn:pion_mass_shift_asymp}, which has
the functional form of our exponential ansatz \eqref{eqn:exp}. We have
seen that the data can be described quite well by the parameterization
\eqref{eqn:exp}. It is therefore instructive to compare our results
for the parameter $c$ to the constant
\begin{equation}
C=\frac{3}{4(2\pi)^{3/2}}\frac{\mps^{3/2}}{(Z_A^{-1}\fps)^2}
\end{equation}
(\cf\ Eq.~\eqref{eqn:pion_mass_shift_asymp}), where at each $(\beta,\kappa)$
we take $\mps$ and $Z_A^{-1}\Fps$ from the largest available lattice,
respectively (see Table~\ref{tab:latt_masses}). For our simulation points
$(\beta,\kappa)=(5.6,0.1575)$, $(5.6,0.158)$ and $(5.32144,0.1665)$ we have
$C/\mathrm{GeV}^{-1/2}=1.089(61)$, $0.745(98)$ and $0.79(12)$,
respectively. (Using $\mpi=137$~MeV and $\fpi=92.4$~MeV the natural value is
$C=0.283\,\mathrm{GeV}^{-1/2}$.) Comparing the first two of these values to
the results for $c$ in Tables~\ref{tab:ldep_fit_params_56_1575} and
\ref{tab:ldep_fit_params_56_1580} (exp,PS) we see that the relative factor
between $c$ and $C$ is $O(10)$ assuming that $Z_A=O(1)$. The discrepancy is
generally larger for smaller values of the left fit boundary, $L_1$, but
decreases for increasing $L_1$.

The large differences between the coefficients $c$ from the fits to our pion
data and $C$ from the L\"uscher formula (with LO ChPT input) reflect the fact
that not all of our data sets for the different volumes at
$(\beta,\kappa)=(5.6,0.1575)$ and $(5.6,0.158)$ comply with the conditions
under which the application of this formula is justified. Recall that these
conditions are: (i)~sufficiently large lattice volumes (because the L\"uscher
formula corresponds to the leading term in a large $L$ expansion), and
(ii)~small pion masses (because we take the pion scattering amplitude from
chiral perturbation theory).

Quite recently, the finite-size shift of the pion mass has been
determined using L\"uscher's formula with the $\pi\pi$ forward
scattering amplitude taken from two-flavor chiral perturbation theory
up to NNLO in the chiral expansion \cite{Colangelo:2003hf}. These
results have then been compared to the leading order chiral expression
for the pion mass in finite volume (including the large-$L$ suppressed
terms neglected by the L\"uscher formula) in order to estimate the
effect of subleading terms in the large $L$ expansion. Both aspects of
this investigation rely on chiral perturbation theory as an expansion
in the pion mass $\mpi$ and the particle momenta $p$, both of which
have to be small compared to the chiral symmetry breaking scale that
is usually identified with $4\pi\fpi$. The conditions of applicability
thus read
\begin{equation}
\label{eqn:mpi_cond}
\frac{\mpi}{4\pi\fpi}\ll 1
\end{equation}
and
\begin{equation}
\frac{p}{4\pi\fpi}\ll 1.
\end{equation}
In a periodic finite box of size $L$, where the particles' momenta can only
take discrete values $p_k=2\pi n_k/L$ with $n_k\in\mathbb{Z}$, the second
condition directly translates into a bound on the box size,
\begin{equation}
\label{eqn:L_cond}
L\gg \frac{1}{2\fpi} \sim 1 \mathrm{fm},
\end{equation}
where we have used the physical value of $\fpi$. (As shown in
Ref.~\cite{Colangelo:2003hf} the pion mass dependence of $\fpi$ predicted by
ChPT at NNLO is rather mild.)  While \textit{a priori} it is not clear what
the practical significance of the relations \eqref{eqn:mpi_cond} and
\eqref{eqn:L_cond} is, we can identify on the basis of Tables~\ref{tab:delta}
and \ref{tab:phys_masses} those data sets that stand the greatest chance of
meeting these conditions. From Table~\ref{tab:delta} one can see that the
ratios $\Mps(L)/(4\pi Z_A^{-1}\Fps(L))$ for all simulated volumes at
$(\beta,\kappa)=(5.32144,0.1665)$ are relatively small and compatible with
each other. The corresponding ratio at $(\beta,\kappa)=(5.6,0.158)$ is also
relatively small for $L=\Lmax$ (and, moreover, comparable to the numbers at
$(5.32144,0.1665)$), but the value for the second largest lattice is already
significantly larger. Considering only the largest lattice, respectively,
$\Mps/(4\pi Z_A^{-1}\Fps)$ is largest at $(\beta,\kappa)=(5.6,0.1575)$, but
here the value at $L=16$ is still consistent with the one at $\Lmax=24$. In
view of this and recalling the relative finite-size mass shifts $R$
(Table~\ref{tab:delta}) we infer from Table~\ref{tab:phys_masses} that for
$\mpi=643$~MeV and $\mpi=490$~MeV we can trust ChPT at most on the largest
volumes with $L\approx 2$~fm (and possibly the $16^3$ lattice with $L\approx
1.4$~fm at $\mpi=643$~MeV), while the lattices with $L<1.4$~fm are most
probably too small. At $(\beta,\kappa)=(5.32144,0.1665)$, on the other hand,
where $\mpi=419$~MeV, all lattices are larger than 1.5~fm due to the
relatively large lattice spacing, and hence appear large enough for ChPT to be
applicable.

\begin{figure}
\begin{center}
\includegraphics*[width=\columnwidth]{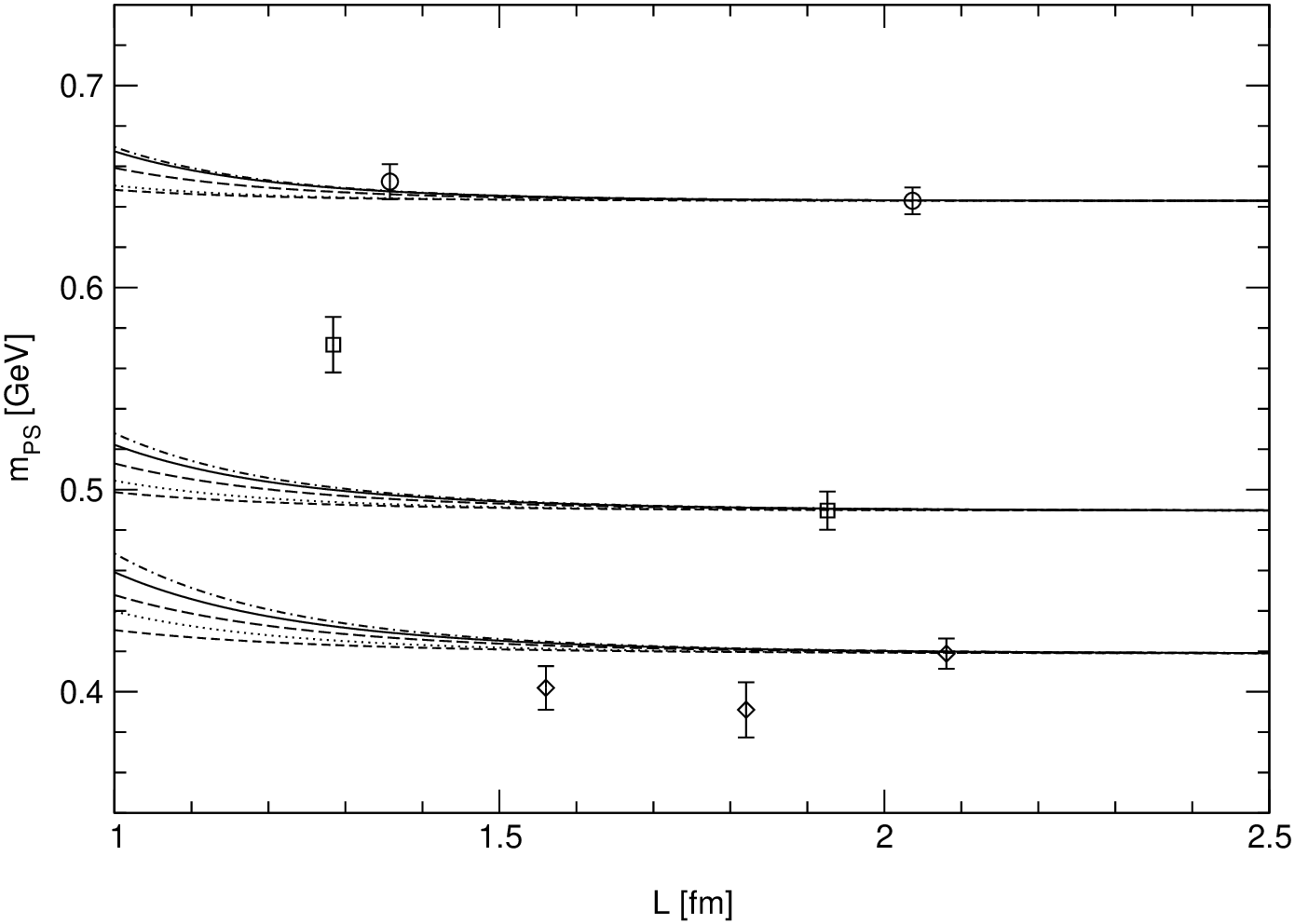}
\end{center}
\caption{\label{fig:ldep_pion}Volume dependence of our pion masses in
  the regime $L\gtrsim 1.3$~fm. The circles, squares and diamonds
  represent our data at $(\beta,\kappa)=(5.6,0.1575)$
  ($\mpi=643$~MeV), $(\beta,\kappa)=(5.6,0.158)$ ($\mpi=490$~MeV) and
  $(\beta,\kappa)=(5.32144,0.1665)$ ($\mpi=419$~MeV), respectively.
  The curves correspond to L\"uscher's formula with input from ChPT at
  LO (dashed), NLO (long-dashed) and NNLO (solid). The dotted curves
  show the full LO chiral expression. The dash-dotted curve is the
  full LO result shifted by the difference between the NNLO and the LO
  L\"uscher formula.}
\end{figure}

In order to corroborate these findings we checked how our simulated
pion masses $\mps(L)$, for $L\gtrsim 1.3$~fm, relate to the results of
Ref.~\cite{Colangelo:2003hf}. There, the chiral expression for
the amplitude $\Fpipi$ has been written as an expansion in powers of
$\xi$,
\begin{equation}
\label{eqn:fpipi_expansion}
\Fpipi = 16\pi^2 \left[ \xi F_2 + \xi^2 F_4 + \xi^3 F_6 + O(\xi^4)
  \right],
\end{equation}
where the parameter $\xi$ is defined as
\begin{equation}
\label{eqn:xi}
\xi = \left(\frac{\mpi}{4\pi\fpi}\right)^2.
\end{equation}
Inserting the expansion \eqref{eqn:fpipi_expansion} up to $F_4$ or $F_6$ into
L\"uscher's formula \eqref{eqn:luescher_pion} for the pion and using the
chiral expression for the isospin invariant amplitude $A$ of
Ref.~\cite{Bijnens:1995yn}, the leading term in the large-$L$ expansion
is obtained up to NLO and NNLO in the chiral expansion. (Correspondingly,
inserting \eqref{eqn:fpipi_expansion} into \eqref{eqn:luescher_pion} only up
to $F_2$ yields the LO expression \eqref{eqn:pion_mass_shift}.)  In order to
calculate the predicted finite size shift for the pion numerically for our
three different pion masses we need to know the respective numerical value of
the expansion parameter $\xi$. In order to avoid the difficulties associated
with the renormalization of the pion decay constant one can use the analytic
expression for the pion mass dependence of $\fpi$ which is known to NNLO in
ChPT. If we take the pion mass from the largest lattice as a first
approximation to the asymptotic pion mass $\mpi$, respectively, we obtain the
curves displayed in Fig.~\ref{fig:ldep_pion}. The dashed curves correspond to
L\"uscher's formula \eqref{eqn:pion_mass_shift} with $\Fpipi$ from ChPT at
leading order. The long-dashed and solid curves show the NLO and NNLO
predictions, respectively. For comparison, the dotted curves show the full
leading-order chiral expression ($N_f=2$) for the pion mass in finite volume,
given by
\begin{equation}
\label{eqn:GL_mpi}
\mpi(L)=\mpi\left[ 1 + \xi \sum_{n=1}^\infty \,m(n)\,
  \frac{K_1(\sqrt{n}\,\mpi L)}{\sqrt{n}\,\mpi L} \right],
\end{equation}
where the multiplicity $m(n)$ counts the number of integer vectors $\vec{n}$
satisfying $n_1^2+n_2^2+n_3^2=n$ \cite{Colangelo:2003hf}. Since the modified
Bessel function $K_1(x)$ falls off exponentially for large $x$, the sum in
\eqref{eqn:GL_mpi} is rapidly converging. For $n=1$ Eq.~\eqref{eqn:GL_mpi}
corresponds precisely to the LO L\"uscher formula
\eqref{eqn:pion_mass_shift}. Finally, the dash-dotted curve in
Fig.~\ref{fig:ldep_pion} represents the best currently available estimate of
the full finite-size effect, obtained by adding to the L\"uscher formula with
$\Fpipi$ at NNLO the difference between Eq.~\eqref{eqn:GL_mpi} and the
L\"uscher formula with $\Fpipi$ at LO.

The main conclusion we draw from the plot is that for all our three
pion masses and for our lattices with $L\gtrsim 1.3$~fm the
finite-size effects predicted by ChPT are considerably smaller than
our statistical errors. On the largest lattices with $L\simeq 2$~fm
the maximal predicted finite-size correction (corresponding to the
dash-dotted curve in the plot) is about 0.3\% for the lightest pion
and 0.05\% for the heaviest one. This is in accordance with our
presumption that for all practical purposes the finite-size effects in
the pion masses are negligible on our largest lattices. At $L\simeq
1.3$~fm the finite-size shift ranges between 1\% for the heaviest and
about 3\% for the lightest pion, which is of the order of the
statistical uncertainties. For $L\gtrsim 1.3$~fm the differences
between the full one-loop ChPT result and L\"uscher's formula with
$\Fpipi$ at LO are comparably small, indicating that here the use of
L\"uscher's asymptotic formula is indeed justified; the maximal
difference in the relative effects is about 50\% at $L\simeq 1.3$~fm
for the smallest pion mass. By contrast, the difference between the
relative effects predicted by L\"uscher's formula with $\Fpipi$ at
NNLO and LO amounts, for the same lattice size, to a factor of 3.2 for
the lightest and 4.5 for the heaviest pion.

Incidentally, a formula analogous to \eqref{eqn:GL_mpi} exists also for the
pion decay constant. (Recently also an asymptotic formula \textit{\`{a} la}
L\"uscher has been derived for $\fpi$ \cite{Colangelo:2004xr}.) The only
difference is that the relative finite size effect is negative and (for
$N_f=2$) four times larger than that of the pion mass:
\begin{equation}
\label{eqn:GL_fpi}
\fpi(L)=\fpi\left[ 1 - 4\xi \sum_{n=1}^\infty \,m(n)\,
  \frac{K_1(\sqrt{n}\,\mpi L)}{\sqrt{n}\,\mpi L} \right].
\end{equation}
We have already seen that the volume dependence of our pion masses can
be accounted for by chiral perturbation theory on the largest lattices
at most, and there is no reason to believe that this should be
different for the decay constant. But we can at least check whether we
can recover the relative factor of minus four. Without going into the
details we just state here that while we find the finite-size effect
of the pion decay constant indeed to be negative, its magnitude is, on
the smaller lattices at $(\beta,\kappa)=(5.6,0.1575)$ and
$(5.6,0.158)$, about the same as that of the pion mass shift; on the
second largest volume at $(\beta,\kappa)=(5.6,0.1575)$ the relative
shift in $\fpi(L)$ is about twice as big as the shift in
$\mpi(L)$. Unfortunately at $(\beta,\kappa)=(5.32144,0.1665)$,
corresponding to our smallest quark mass, we can make no definite
statement.

\subsubsection{\label{sec:nucleon}Nucleon}

\begin{figure}
\includegraphics*[width=\columnwidth]{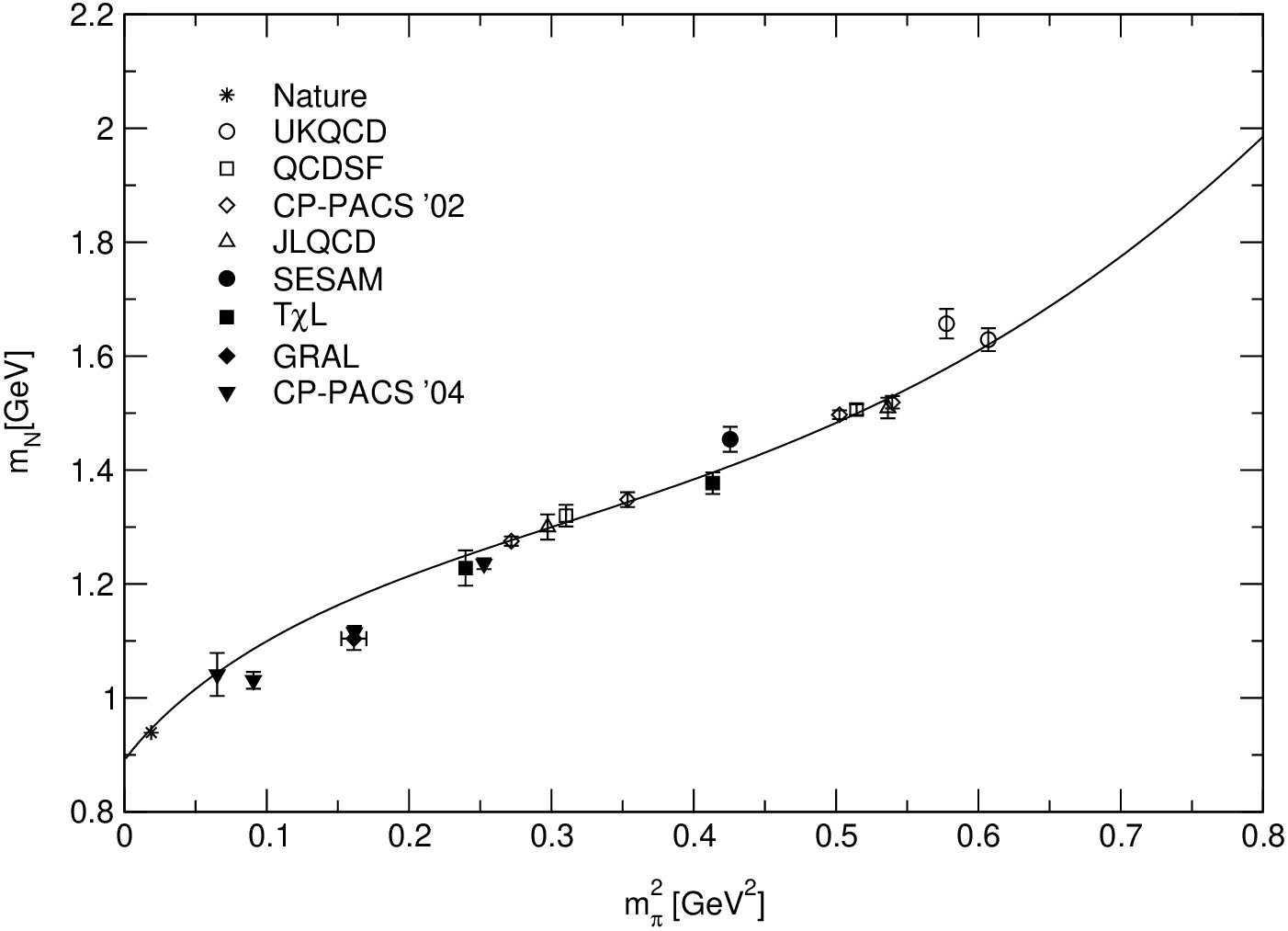}
\caption{\label{fig:mNmpisq_dep}Nucleon mass data from various collaborations
  as a function of $\mpi^2\propto m_q$, including our data. The curve
  corresponds to a fit of the data represented by the open symbols to
  Eq.~\eqref{eqn:mNmpisq}. These data points are from simulations on
  relatively large and fine lattices. Note that the fit result is consistent
  with the physical pion and nucleon masses (indicated by the star).}
\end{figure}

Regarding the nucleon mass, replacing the simple exponential \eqref{eqn:exp}
by an ansatz corresponding more closely to L\"uscher's nucleon mass shift
formula of Ref.~\cite{Luscher:1983rk} might be considered as the natural next
step towards a better description of the volume dependence. (Although, as we
have seen, the ansatz \eqref{eqn:exp} describes the data already quite well.)
But since L\"uscher's nucleon formula can be seen as a special case of the
formula \eqref{eqn:delta_Op4}, let us instead confront our data for the
nucleon mass directly with the formulae \eqref{eqn:delta_Op3} and
\eqref{eqn:delta_Op4}. Following Ref.~\cite{AliKhan:2003cu} we fix $g_A$ and
$\fpi$ to the physical values $g_A=1.267$, $\fpi=92.4$~MeV, and set the
couplings $c_2$ and $c_3$ to $c_2=3.2\,\mathrm{GeV}^{-1}$,
$c_3=-3.4\,\mathrm{GeV}^{-1}$. The remaining parameters $m_0$, $c_1$ and
$e_1^r(\lambda)$ (where the renormalization scale $\lambda$ is chosen to be
1~GeV) are taken from a fit of data from various unquenched simulations with
\begin{equation}
\label{eqn:conditions}
a<\text{0.15~fm}, \quad \mpi L>5 \quad \text{and} \quad \mpi<\text{800~MeV}
\end{equation}
to Eq.~\eqref{eqn:mNmpisq}. In Ref.~\cite{AliKhan:2003cu}, data from the QCDSF
\cite{AliKhan:2003cu}, UKQCD \cite{Allton:2001sk}, CP-PACS
\cite{AliKhan:2001tx} and JLQCD \cite{Aoki:2002uc} collaborations have been
used. These data are plotted in Fig.~\ref{fig:mNmpisq_dep} (open symbols),
complemented by the results from our largest lattices, namely the \txl\
results at $(\beta,\kappa,L)=(5.6,0.1575,24), (5.6,0.158,24)$ and the GRAL
result at $(5.32144,0.1665,16)$. We also include the SESAM result at
$(5.6,0.1575,16)$ in the plot, and recent results from CP-PACS for small quark
masses but from quite coarse lattices \cite{Namekawa:2004bi} (solid
symbols). Although the conditions~\eqref{eqn:conditions} are to some extent
arbitrary we stick to them for definiteness. Consequently we refrain from
repeating the fits of Ref.~\cite{AliKhan:2003cu} with our or the new CP-PACS
data, because only the \txl\ point at $(\beta,\kappa,L)=(5.6,0.1575,24)$ meets
all of the requirements in \eqref{eqn:conditions}. Instead we quote the result
of fit~1 in Ref.~\cite{AliKhan:2003cu} where $m_0=0.89(6)$~GeV,
$c_1=-0.93(5)\,\mathrm{GeV}^{-1}$ and
$e_1^r(\lambda=1\,\mathrm{GeV})=2.8(4)\,\mathrm{GeV}^{-3}$, consistent with
phenomenology. The corresponding curve is represented by the solid line in
Fig.~\ref{fig:mNmpisq_dep}. The fact that the \txl\ point at
$(\beta,\kappa,L)=(5.6,0.1575,24)$ lies close to the curve without having been
included into the fit hints to a small $O(a)$ effect at this point.

Note that we use the standard Wilson plaquette and quark action with errors at
$O(a)$, whereas the data from the other collaborations have all been generated
with $O(a)$-improved actions.

 The other \txl\ point at $(\beta,\kappa,L)=(5.6,0.158,24)$, corresponding to
a smaller pion mass, lies somewhat below the curve. Correcting it for the
presumed finite-size effects in the pion and the nucleon mass would shift it
even slightly further away from the curve (recall that in this regime of
larger $L$ the finite-size effect is bigger for the nucleon than for the
pion). The SESAM point at $(\beta,\kappa,L)=(5.6,0.1575,16)$ illustrates how
finite-size effects appear in such a plot. Correcting it for the finite-size
effects in the pion and the nucleon masses (see Table~\ref{tab:delta}) would
shift it to the lower left, towards the corresponding \txl\ point with
$L=24$. The data points from our largest lattices generally tend to lie
somewhat below the curve, and this is also true for the GRAL point with
$(\beta,\kappa)=(5.32144,0.1665)$. In view of the fluctuations in $\mpi(L)$ at
this parameter set we plot in Fig.~\ref{fig:mNmpisq_dep} the mean of the
respective pion masses at $L=12,14,16$, with a corresponding error bar along
the $\mpi^2$ axis. Even with this uncertainty taken into account the deviation
of the GRAL point from the fit curve is significant. Considering the
relatively low cut-off of only about 1.5~GeV at this point (to be compared
to a nucleon mass of 1.1~GeV) discretization errors might be responsible for
the deviation. In case of the \txl\ data cut-off effects are expected to be
less important, due to the smaller lattice spacings in these simulations.

\begin{figure}
\includegraphics*[width=\columnwidth]{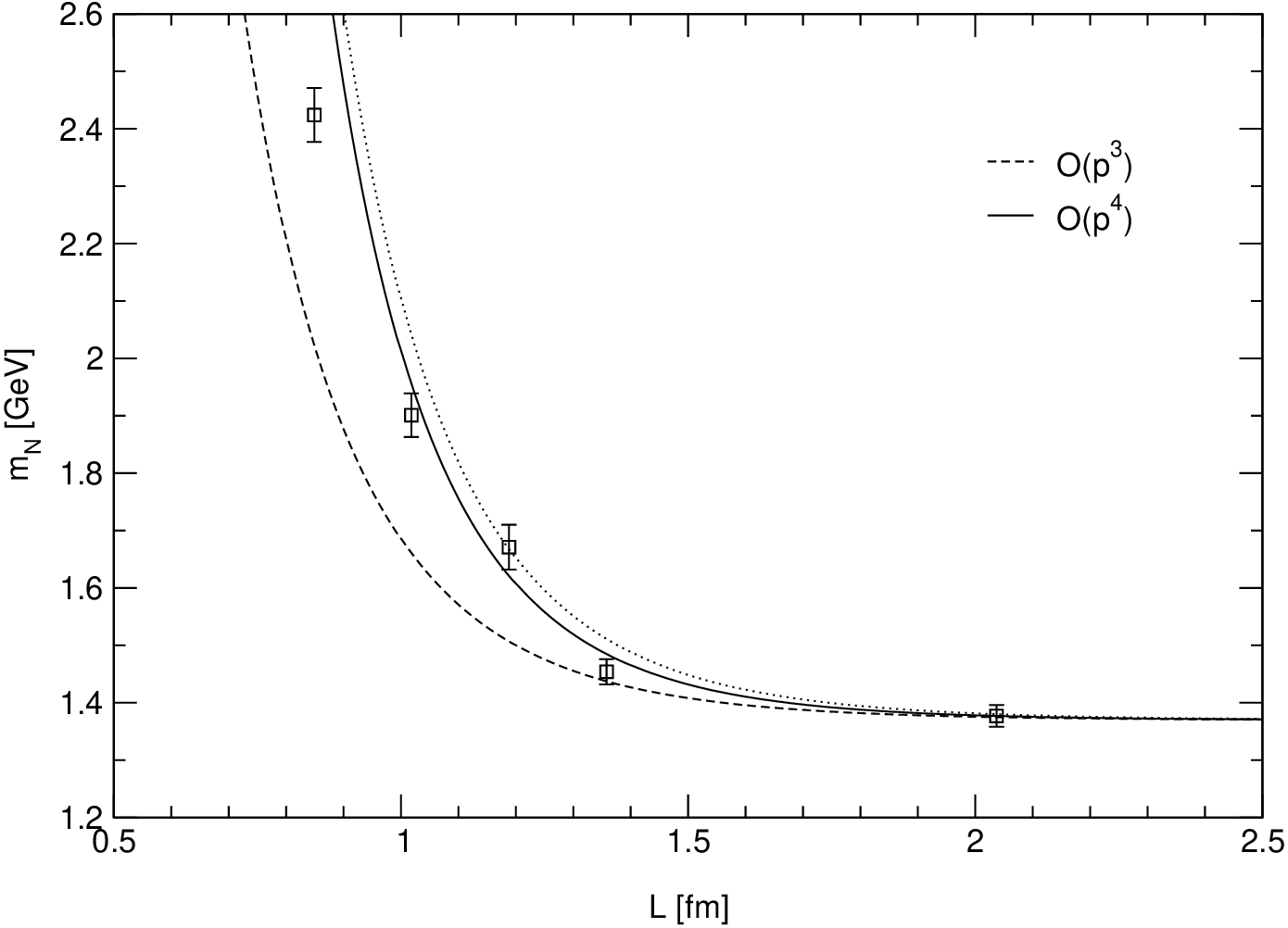}
\caption{\label{fig:nucleon_56_1575}Volume dependence of the
  nucleon mass for $\mpi=643$~MeV. The dashed curve represents the
  $O(p^3)$ term only, while the solid curve also includes the $O(p^4)$
  contribution. The dotted curve results if the pion mass is reduced
  by 10\% in the $O(p^4)$ formula.}
\end{figure}

\begin{figure}
\includegraphics*[width=\columnwidth]{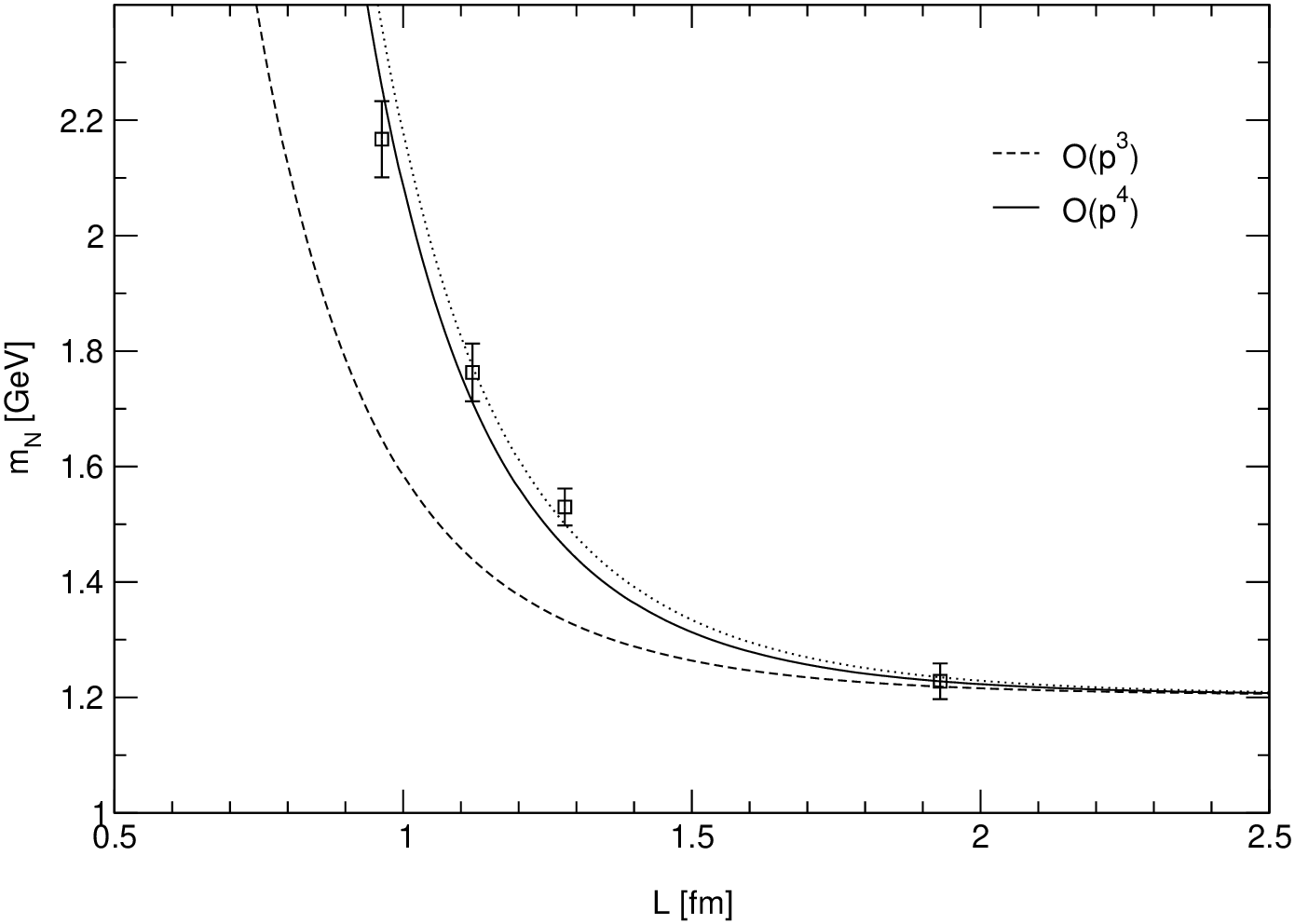}
\caption{\label{fig:nucleon_56_1580} Same as
  Fig.~\ref{fig:nucleon_56_1575} for $\mpi=490$~MeV.}
\end{figure}

\begin{figure}
\includegraphics*[width=\columnwidth]{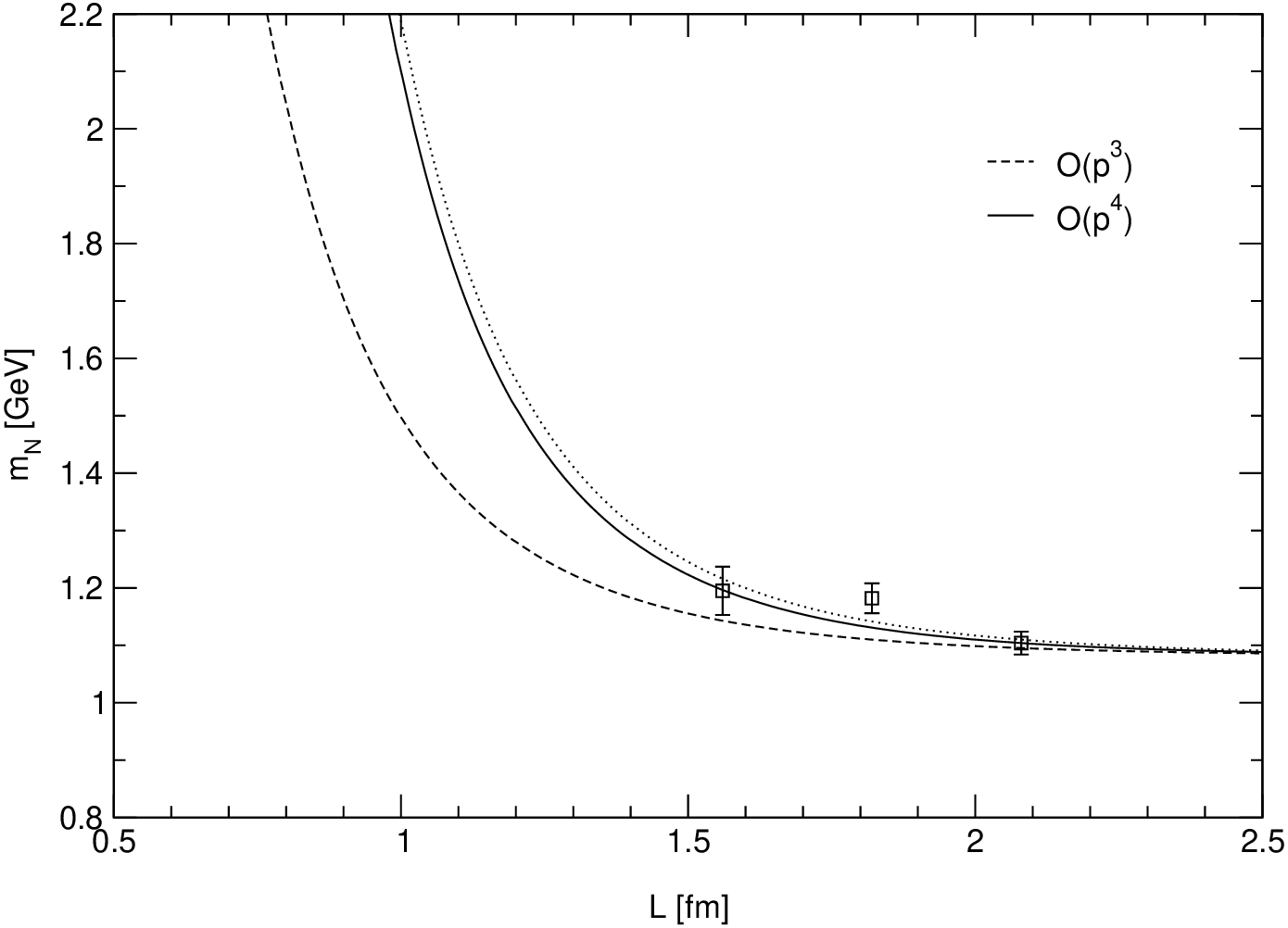}
\caption{\label{fig:nucleon_532144_1665} Same as
  Fig.~\ref{fig:nucleon_56_1575} for $\mpi=419$~MeV.}
\end{figure}

Using the parameters corresponding to the solid curve in
Fig.~\ref{fig:mNmpisq_dep} we can evaluate the finite-size formulae
\eqref{eqn:delta_Op3} and \eqref{eqn:delta_Op4} and compare the
results to our data. Our three sets of simulations with different
lattice sizes correspond to pion masses of approximately 643, 490
and 419~MeV. Note that the latter two masses are lighter than the
lightest of the pion masses investigated in
Ref.~\cite{AliKhan:2003cu} (732, 717 and 545~MeV). The
curves in
Figs.~\ref{fig:nucleon_56_1575}--\ref{fig:nucleon_532144_1665} have
been computed from equations \eqref{eqn:delta_Op3} and
\eqref{eqn:delta_Op4} with no free parameters. Like in
Ref.~\cite{AliKhan:2003cu} the solid curves correspond to the
$O(p^4)$ prediction
\begin{equation}
\label{eqn:delta_tot_2}
m_N(L) = m_N + \Delta_a(L) + \Delta_b(L),
\end{equation}
where $m_N$ has been determined such that the calculated value $m_N(\Lmax)$
equals the simulated mass from the largest lattice with $L\!=\!\Lmax$,
respectively. Correspondingly, for the pion masses $\mpi$ we also take the
simulated value from the largest lattices. For the dashed curve, corresponding
to the $O(p^3)$ prediction, the $O(p^4)$ contribution from $\Delta_b$ in
\eqref{eqn:delta_tot_2} has been omitted, while $m_N$ has been left
unchanged. For all our pion masses we find a surprisingly good overall
description of our data by the $O(p^4)$ prediction even down to lattice sizes
of about 1~fm. Replacing $\mpi$ from the largest, $L\!=\!16$ lattice at
$(\beta,\kappa)=(5.32144,0.1665)$ by the mean of the pion masses from the
$L=12,14,16$ lattices (as we did in Fig.~\ref{fig:mNmpisq_dep}) does not lead
to a significant difference in the resulting curve. Since both the statistical
and the theoretical errors of the simulated $m_N(L)$ are smallest for the
largest lattice, we consider the $O(p^4)$ finite-size corrected nucleon mass
\begin{equation}
\label{eqn:fse_nucleon}
\tilde{m}_N(L) = m_N(L) - \Delta_a(L) - \Delta_b(L),
\end{equation}
taken at $L\!=\!\Lmax$, as our best estimate of the asymptotic nucleon
mass. Table~\ref{tab:nucleon} shows the predicted infinite-volume
masses for our simulations. The last column gives the relative mass
shift on the largest lattice, respectively. Just as it was the case
for the pion, the finite size effect in the nucleon at
$(\beta,\kappa,L)=(5.6,0.1575,24)$ is considerably smaller than the
statistical uncertainty. At $(\beta,\kappa,L)=(5.6,0.158,24)$ and
$(5.32144,0.1665,16)$, on the other hand, the finite-size effects
according to Eq.~\eqref{eqn:fse_nucleon} amount to about 2\% of the
respective asymptotic mass, which is comparable to the statistical
errors.

\begin{table*}
 \caption{\label{tab:nucleon}Finite-size shifts of the nucleon masses
  on the largest lattices as inferred from
  Eq.~\eqref{eqn:delta_tot_2}. $\Delta =
  (m_N-\tilde{m}_N)/\tilde{m}_N$ is the relative deviation of the
  Monte Carlo value $m_N$ from the shifted mass
  $\tilde{m}_N=m_N-\Delta_a-\Delta_b$ (all values to be taken at
  $\Lmax$). We consider $\tilde{m}_N(\Lmax)$ as the best estimate of
  the true asymptotic mass.}
 \begin{ruledtabular}
  \begin{tabular}{lllccc}
   \cc{$\beta$} & \cc{$\kappa$} & \cc{$\mps\,[\mathrm{GeV}]$} &
   \cc{$\tilde{m}_N(\Lmax)\,[\mathrm{GeV}]$} & 
   \cc{$m_N(\Lmax)\,[\mathrm{GeV}]$} & \cc{$\Delta$} \\
   \hline
   5.6      & 0.1575 & 0.6429(67) & 1.370(19) & 1.377(19) & 0.53\% \\
            & 0.1580 & 0.4896(94) & 1.204(31) & 1.228(31) & 2.02\% \\
   5.32144  & 0.1665 & 0.4188(75) & 1.081(20) & 1.104(20) & 2.08\% \\
  \end{tabular}
 \end{ruledtabular}
\end{table*}

Compared to a fit-based extrapolation the advantage of a formula without free
parameters is of course that one can directly calculate the amount by which
one has to shift the nucleon mass in order to compensate for the finite-size
effect associated with a given volume, and that one has control over the
error. In practice, however, a remaining caveat is that the infinite-volume
pion mass must be known. If one is working in a parameter regime where the
finite-size effect in the pion mass is small (of the order of a few percent)
one can apply the results of Ref.~\cite{Colangelo:2003hf} to obtain an
estimate of the true asymptotic mass. If this is unclear, but data from
several (more than two) different volumes are available, one might still
revert to an exponential fit and extrapolate. Since we have seen that such a
``naive'' extrapolation systematically underestimates the true infinite-volume
pion mass we illustrate, as an example, the impact of a by 10\% smaller pion
mass by the dotted curves in
Figures~\ref{fig:nucleon_56_1575}--\ref{fig:nucleon_532144_1665}. Although
relative to the very nucleon mass shift the systematic error associated with
the uncertainty in the pion mass grows with $L$, its absolute value becomes
less and less significant compared to the statistical errors of the data. On
the one hand this means that (assuming the formula to exactly reproduce the
volume dependence of the data and the statistical uncertainties to be all of
comparable size) in order to predict the asymptotic nucleon mass correctly
(within the statistical errors) one needs to know $\mpi$ the more accurately
the smaller the physical size of the largest available lattice. If, on the
other hand, $L$ is sufficiently large so that one can reliably extrapolate the
pion mass, the asymptotic nucleon mass can be determined quite accurately,
already on the basis of a single lattice.

\subsection{\label{sec:loops}Spatial Polyakov-type loops}
 
At our two larger quark masses we have observed in all considered
quantities (pion, rho and nucleon masses, pion decay constant) a
drastic increase of the finite-size shifts below a lattice size of
approximately 1.5~fm. But above this size the finite-size effects
are relatively small, also at our smallest quark mass. As we will now
show, this kind of transition behavior is reflected in the behavior
of spatial Polyakov-type loops.

In the absence of dynamical quarks, \ie\ in quenched QCD, the
expectation value of the Polyakov loop, which is defined as
\begin{equation}
  \label{eqn:polyakov_loop}
  \ev{P} = \left\langle\frac{1}{L^3} \sum_{\xvec} \mathrm{Tr}
  \prod_{\tau=1}^T U_4(\xvec,\tau)\right\rangle,
\end{equation}
is zero in the confined phase, while in the deconfined phase
$\av{P}\neq 0$. Therefore in pure gauge theory $\av{P}$ is an order
parameter for the deconfining phase transition. This is due to the
global $Z(3)$ symmetry of the pure $\su{3}$ gauge theory which is
spontaneously broken at the phase transition. In full QCD the Polyakov
loop is not an order parameter because the $Z(3)$ symmetry of the
gluonic action is explicitly broken by the quark action, so that
$\av{P}$ is not exactly zero in the hadronic phase. In our simulations
it is not the time extent $T$ which is varied, but the spatial lattice
size $L$. Due to the space-time symmetry of the Euclidean metric,
however, similar considerations also apply to Polyakov-type loops in
the spatial directions (also known as \emph{Wilson lines}). Let us,
for definiteness, consider the mean Wilson line in $z$-direction,
which is defined configuration-wise as
\begin{equation}
  \label{eqn:wilson_line}
  P_z = \frac{1}{L^2T} \sum_{x_1,x_2,\tau} \mathrm{Tr}
  \prod_{x_3=1}^L U_3(x_1,x_2,x_3,\tau).
\end{equation}

\begin{table}
 \caption{\label{tab:wilson_lines}Ensemble averages of the real and
  imaginary parts of the mean spatial Polyakov loop (in
  $z$-direction).}
 \begin{ruledtabular}
  \begin{tabular}{lllll}
   \cc{$\beta$} & \cc{$\kappa$} & \cc{$L^3T$} & \cc{$\ev{\mathrm{Re}\,P_z}$} & \cc{$\ev{\mathrm{Im}\,P_z}$}\\
   \hline
   5.32144 & 0.1665 & $12^3 32$ &           -0.00056433(3)  &           -0.0003573(2)  \\
           &        & $14^3 32$ &           -0.0000179(2)   &           -0.00003209(4) \\
           &        & $16^3 32$ & \phantom{-}0.00012251(3)  &           -0.00021856(4) \\
   5.6     & 0.1575 & $10^3 32$ &           -0.0131032(3)   &           -0.006341(8)   \\
           &        & $12^3 32$ &           -0.0029148(3)   & \phantom{-}0.0012405(6)  \\
           &        & $14^3 32$ &           -0.00083854(3)  & \phantom{-}0.00008748(3) \\
           &        & $16^3 32$ & \phantom{-}0.00020948(3)  & \phantom{-}0.00004254(5) \\
           &        & $24^3 40$ &           -0.000078415(8) &           -0.00009462(1) \\
           & 0.1580 & $12^3 32$ &           -0.003798(2)    & \phantom{-}0.005014(3)   \\
           &        & $14^3 32$ &           -0.00135725(6)  & \phantom{-}0.0003214(3)  \\
           &        & $16^3 32$ &           -0.00037040(4)  &           -0.00022400(4) \\
           &        & $24^3 40$ &           -0.00010810(2)  & \phantom{-}0.00012503(1) \\
  \end{tabular}
 \end{ruledtabular}
\end{table}

As can be seen from Table~\ref{tab:wilson_lines}, the expectation
values $\av{P_z}$ for all GRAL simulations are indeed significantly
different from zero, even on the largest lattices. While for the
larger lattices the deviation from zero is relatively small it becomes
more pronounced as the lattices shrink. $\av{\mathrm{Re}\,P_z}$ in
particular takes increasingly negative values towards the smaller
volumes. This can be understood by looking at the distribution of
$P_z$.

\begin{figure}
\includegraphics*[width=\columnwidth]{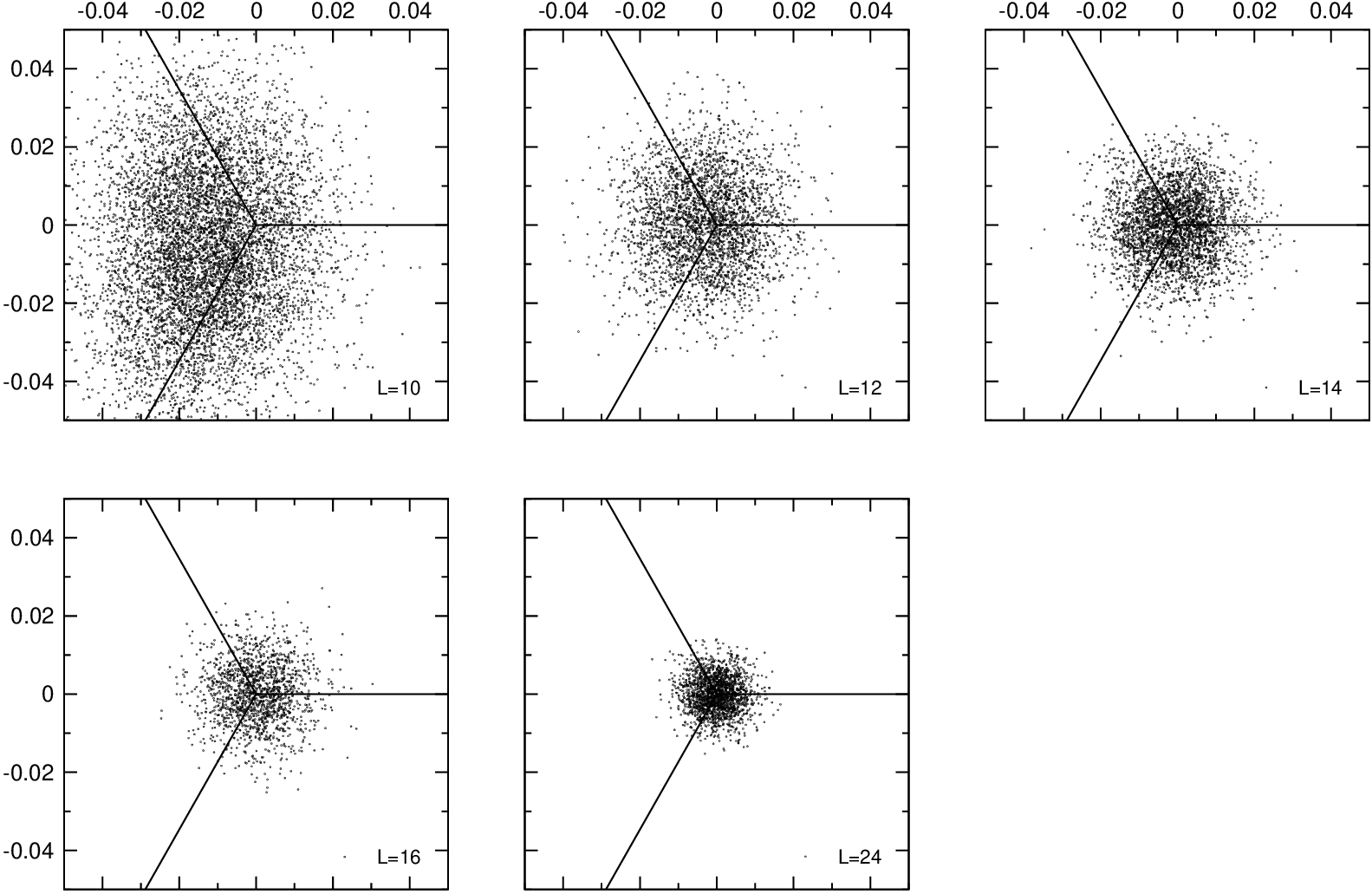}
\caption{\label{fig:polyakov_B5.6_K0.1575}Distribution in the complex
  plane of the spatial Polyakov loop $P_z$ for the different lattices
  simulated at $(\beta,\kappa)=(5.6,0.1575)$. The lines indicate the
  three $Z(3)$ directions $1$, $e^{2\pi i/3}$ and $e^{4\pi i/3}$.}
\end{figure}

Fig.~\ref{fig:polyakov_B5.6_K0.1575} shows the distribution in the
complex plane of $P_z$ for the lattice volumes simulated at
$(\beta,\kappa)=(5.6,0.1575)$. The lines in the plots indicate the
three $Z(3)$ directions $1$, $e^{2\pi i/3}$ and $e^{4\pi i/3}$. Apart
from the $L$-dependent fluctuations we observe for the larger lattices
an approximately point-symmetric accumulation of the Wilson line
around zero. This is reflected in the smallness of $\abs{\av{P_z}}$
and the corresponding statistical errors at large $L$
(Table~\ref{tab:wilson_lines}). The situation is somewhat different
for the smallest, $10^3$ lattice, where the distribution of Wilson
lines is visibly shifted towards the $Z(3)$ directions $e^{2\pi i/3}$
and $e^{4\pi i/3}$, which leads to the relatively large negative value
of $\av{\mathrm{Re}\,P_z}$.

This shift can be understood \eg\ by means of the 3-d Potts model with
magnetic field that is recovered when the full QCD action is expanded first in
the gauge coupling $\beta$ and then in inverse powers of the sea quark mass
\cite{Antonelli:1994ea}. Introducing the quark action in QCD is then
equivalent to switching on a magnetic field $h$ in the Potts model that breaks
the $Z(3)$ symmetry of the system. Considering the phase of the spatial
Polyakov loop as a spin that can take one of the three possible values $1$,
$e^{2\pi i/3}$ and $e^{4\pi i/3}$, the magnetic field aligns the spins to
preferred directions depending on the sign of $h$: for $h=-\abs{h}$
(corresponding to antiperiodic spatial boundary conditions for sea quarks) the
positive real axis is favored, whereas for $h=+\abs{h}$ (periodic spatial
boundary conditions for sea quarks) the two directions $e^{2\pi i/3}$ and
$e^{4\pi i/3}$ (pointing towards negative values) are preferred. If we recall
that we have used periodic spatial boundary conditions for sea quarks this
explains the plot for $L=10$ in Fig.~\ref{fig:polyakov_B5.6_K0.1575}.

The implications of such a shift for finite-size effects in hadron
masses have been explained in detail \eg\ by Aoki \etal\ in the
context of their comparative study of finite-size effects in quenched
and full QCD simulations \cite{Aoki:1993gi}. We briefly recapitulate
their argument for our choice of boundary conditions: Let us consider
a meson propagator $\Gamma(\tau)$ on a lattice of size $L^3$ with a
sufficiently large time extent $T$. A hopping parameter expansion of
$\Gamma(\tau)$ yields a representation of the meson propagator in
terms of closed valence quark loops $C$ going through the meson source
and sink. If we denote the corresponding link factors $\tr \prod_{l\in
C} U_l$ by $P(C)$ for Polyakov-type loops that wind around the lattice
in a spatial direction, and $W(C)$ for ordinary Wilson-type loops, the
meson propagator can be written as
\begin{equation}
\label{eqn:meson_prop}
-\ev{\Gamma(\tau)} = \sum_C \kval^{L(C)} \ev{W(C)} + \sum_C
 \kval^{L(C)} \sigma_\mathrm{val} \ev{P(C)},
\end{equation}
where $L(C)$ is the length of the respective loop and the sign factor
$\sigma_\mathrm{val}$ is equal to $+1$ for the periodic spatial
boundary conditions used for valence quarks in our simulations. From
the discussion above we know that in the case of periodic spatial
boundary conditions for both sea and valence quarks the contribution
of Polyakov-type loops to the meson propagator \eqref{eqn:meson_prop}
is negative. Since mean values of Wilson-type loops are always
positive, the two contributions in \eqref{eqn:meson_prop} have
opposite sign, which leads to a faster decrease of the correlator and
thus to a larger meson mass.

\begin{figure}
\includegraphics*[width=\columnwidth]{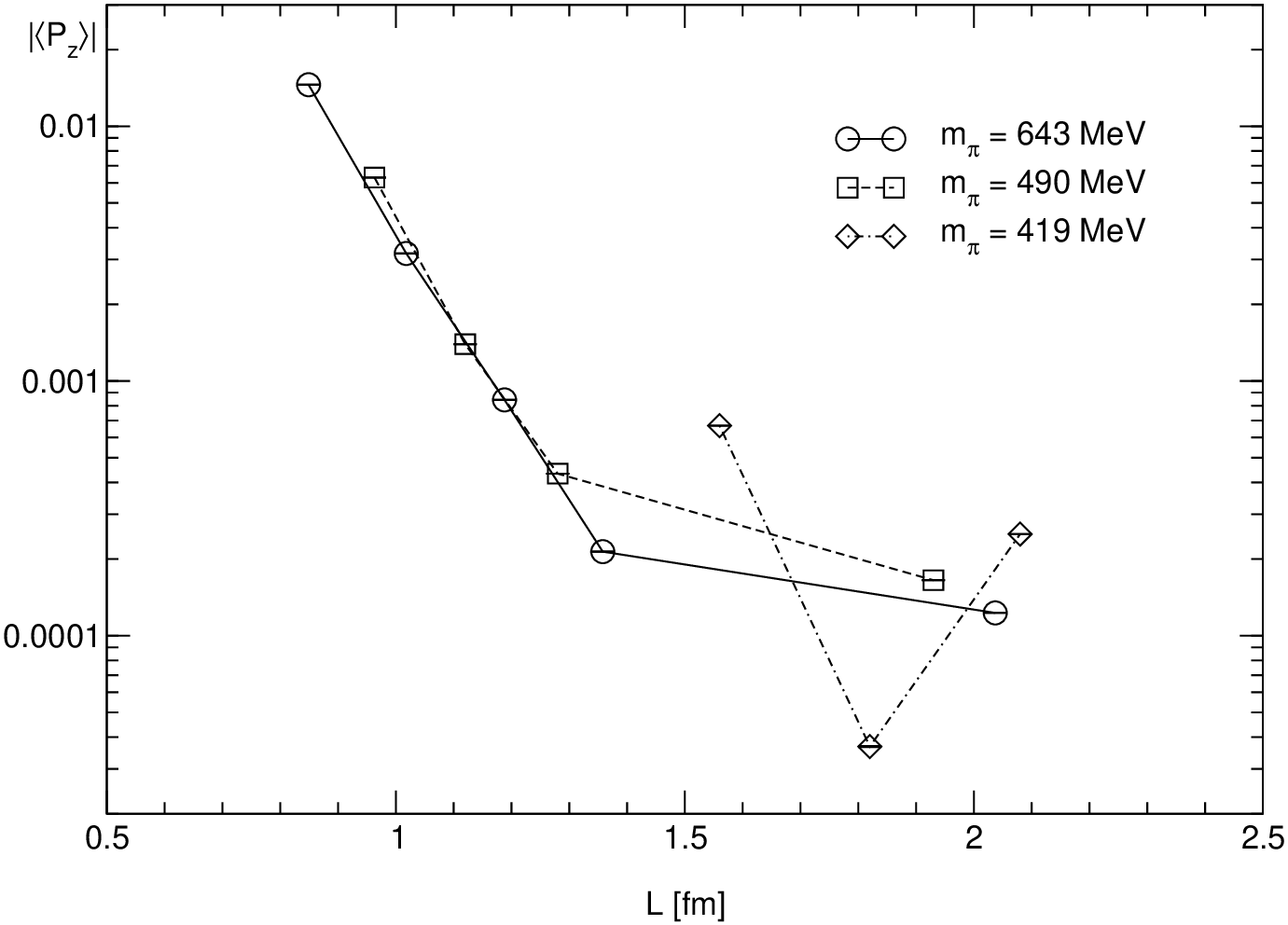}
\caption{\label{fig:wilson_lines_abs}Absolute values of the complex
ensemble averages of the spatial Polyakov loop $P_z$.}
\end{figure}

For fixed sea and valence quark mass this effect grows weaker for increasing
lattice size because the contribution of the Polyakov-type loops
decreases. This can clearly be seen from Fig.~\ref{fig:wilson_lines_abs} for
our larger pion masses. On the other hand, in a fixed lattice volume with
periodic boundary conditions finite-size effects in hadron masses get
increasingly significant both for decreasing sea and valence quark mass. This
has been observed \eg\ in Ref.~\cite{Eicker:diss}, where partially
quenched chiral extrapolations of the pseudoscalar and vector masses were
studied for the various sea and valence quark masses at $\beta=5.5$ and
$5.6$. The sea quark mass dependence of the expectation value of the Wilson
line can also be seen in Table~\ref{tab:wilson_lines} if one compares at
$\beta=5.6$ the value for a given lattice size at $\kappa=0.1575$ with the
corresponding value at $\kappa=0.158$. We find that in the same volume
$\av{\mathrm{Re}\,P_z}$ is more negative for the larger $\kappa$,
corresponding to a smaller quark mass. This leads to a stronger cancellation
of the two terms in \eqref{eqn:meson_prop} and hence to a larger finite-size
effect in the hadron masses.

\begin{figure}
\includegraphics*[width=\columnwidth]{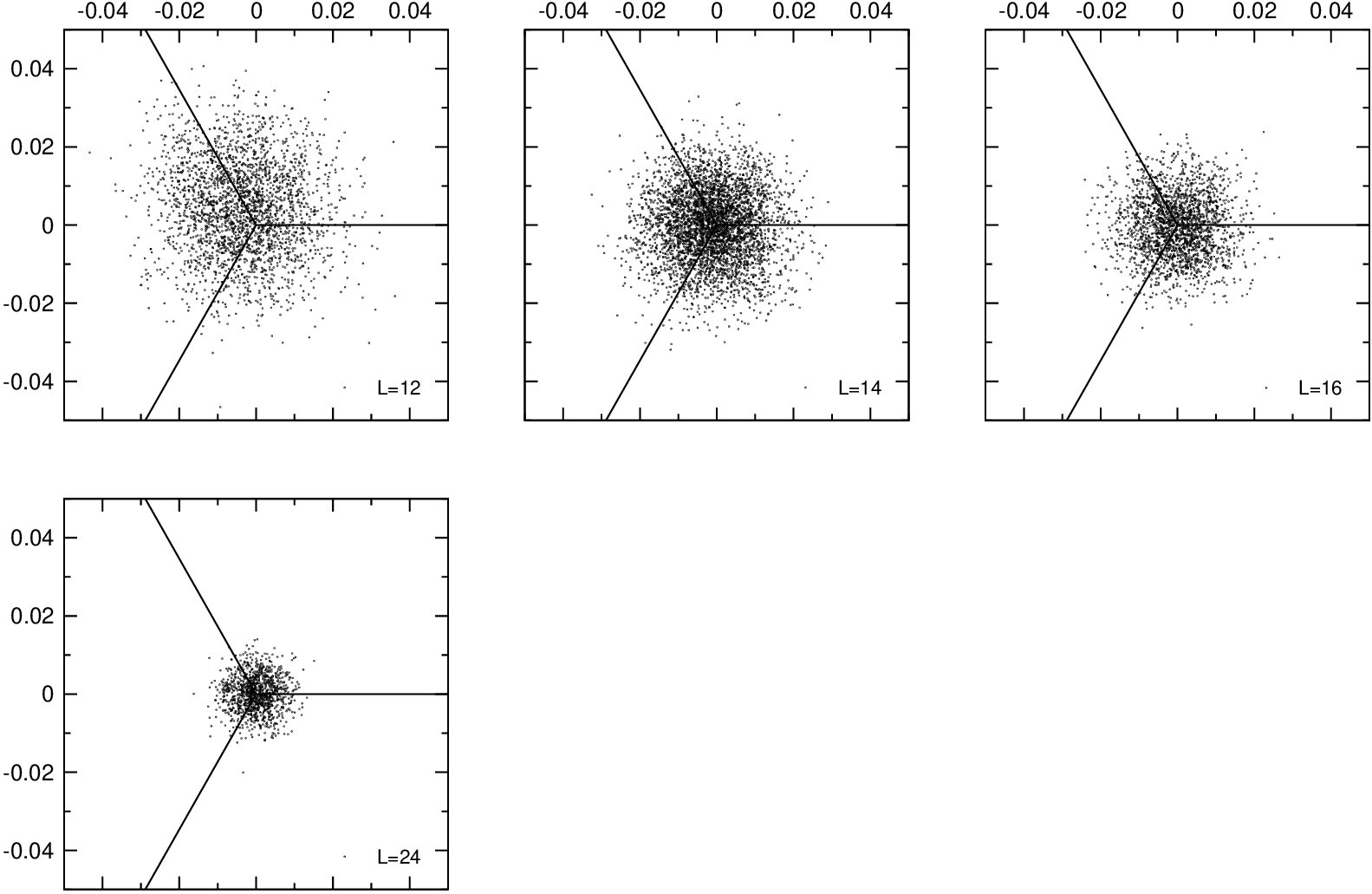}
\caption{\label{fig:polyakov_B5.6_K0.1580} Same as
  Fig.~\ref{fig:polyakov_B5.6_K0.1575} for
  $(\beta,\kappa)=(5.6,0.158)$.}
\end{figure}

\begin{figure}
\includegraphics*[width=\columnwidth]{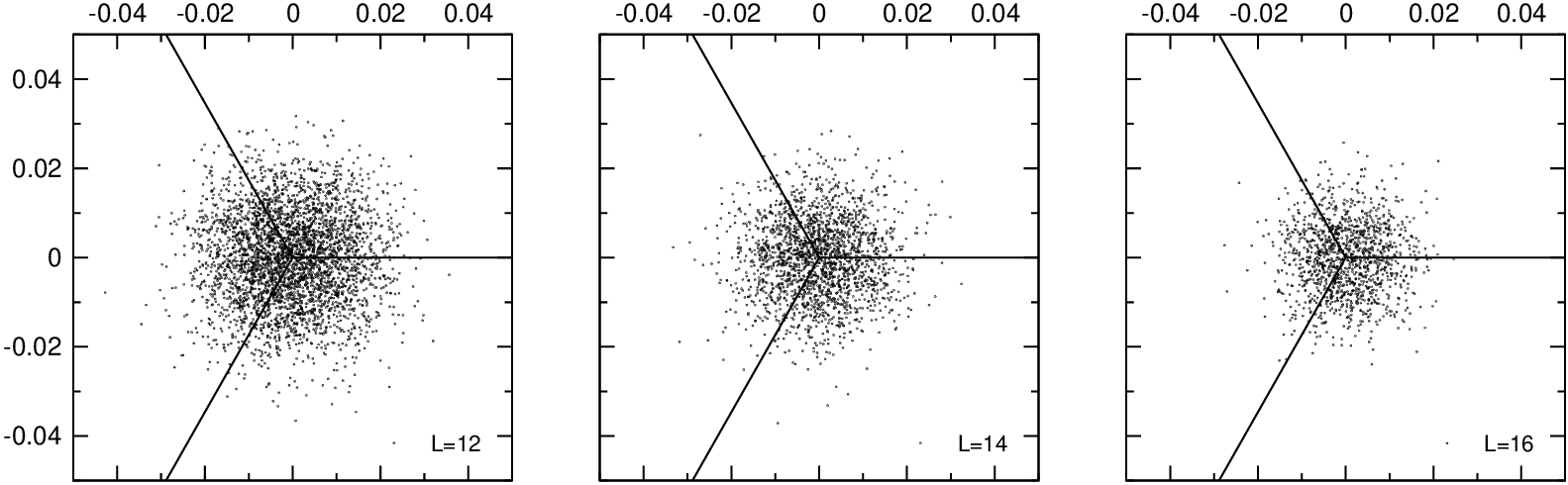}
\caption{\label{fig:polyakov_B5.32144_K0.1665} Same as
  Fig.~\ref{fig:polyakov_B5.6_K0.1575} for
  $(\beta,\kappa)=(5.32144,0.1665)$.}
\end{figure}

For completeness, Figs.~\ref{fig:polyakov_B5.6_K0.1580} and
\ref{fig:polyakov_B5.32144_K0.1665} show the distribution of the
Wilson line for the simulated volumes at $(\beta,\kappa)=(5.6,0.158)$
and $(5.32144,0.1665)$.

\section{\label{sec:cutoff_effects}Discretization Errors}

\begin{figure}
 \includegraphics*[width=\columnwidth]{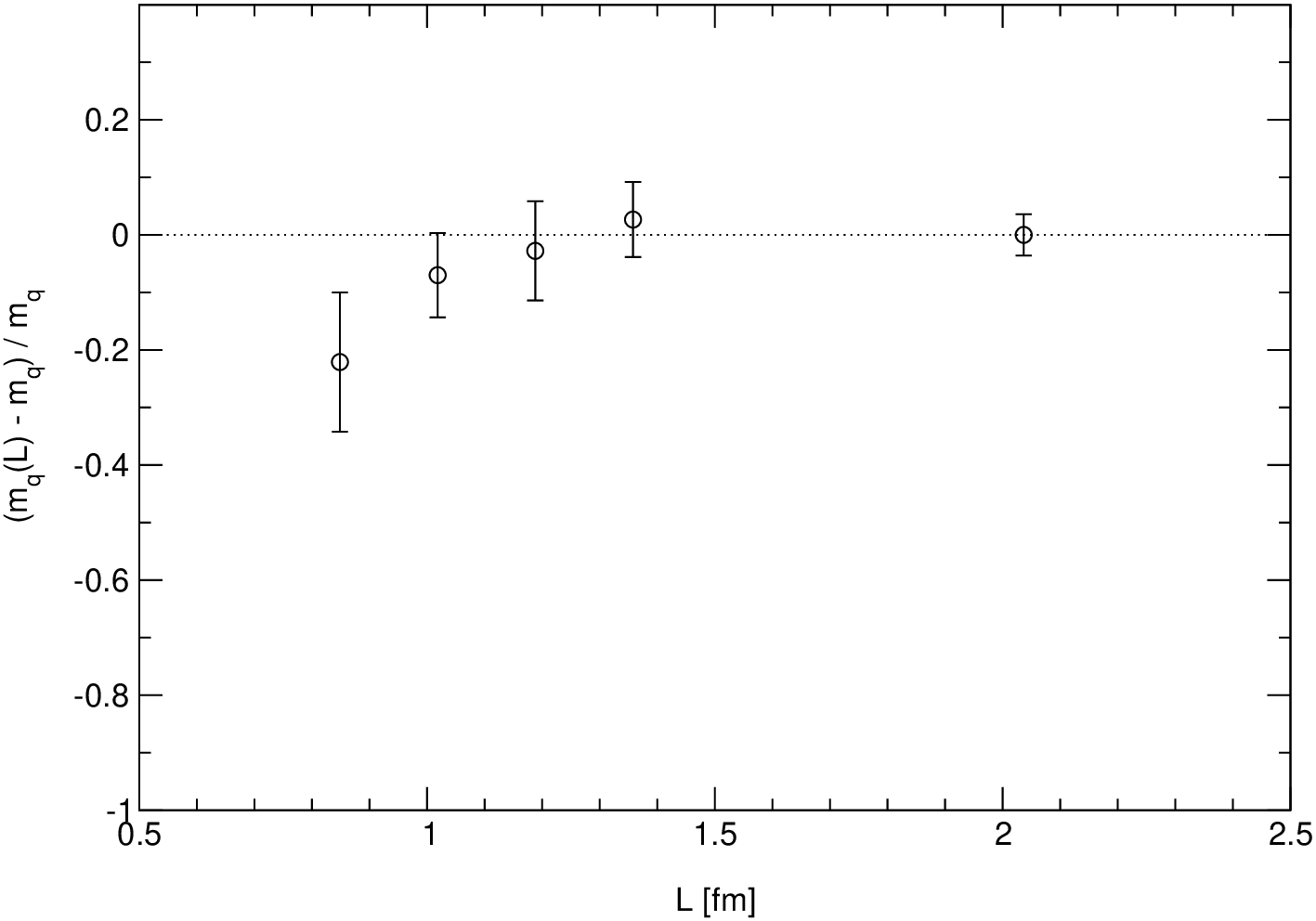}
 \caption{\label{fig:ldep_mq_5.6_1575}Box-size dependence of the
  relative shift in the PCAC quark mass at
  $(\beta,\kappa)=(5.6,0.1575)$.}
\end{figure}

\begin{figure}
 \includegraphics*[width=\columnwidth]{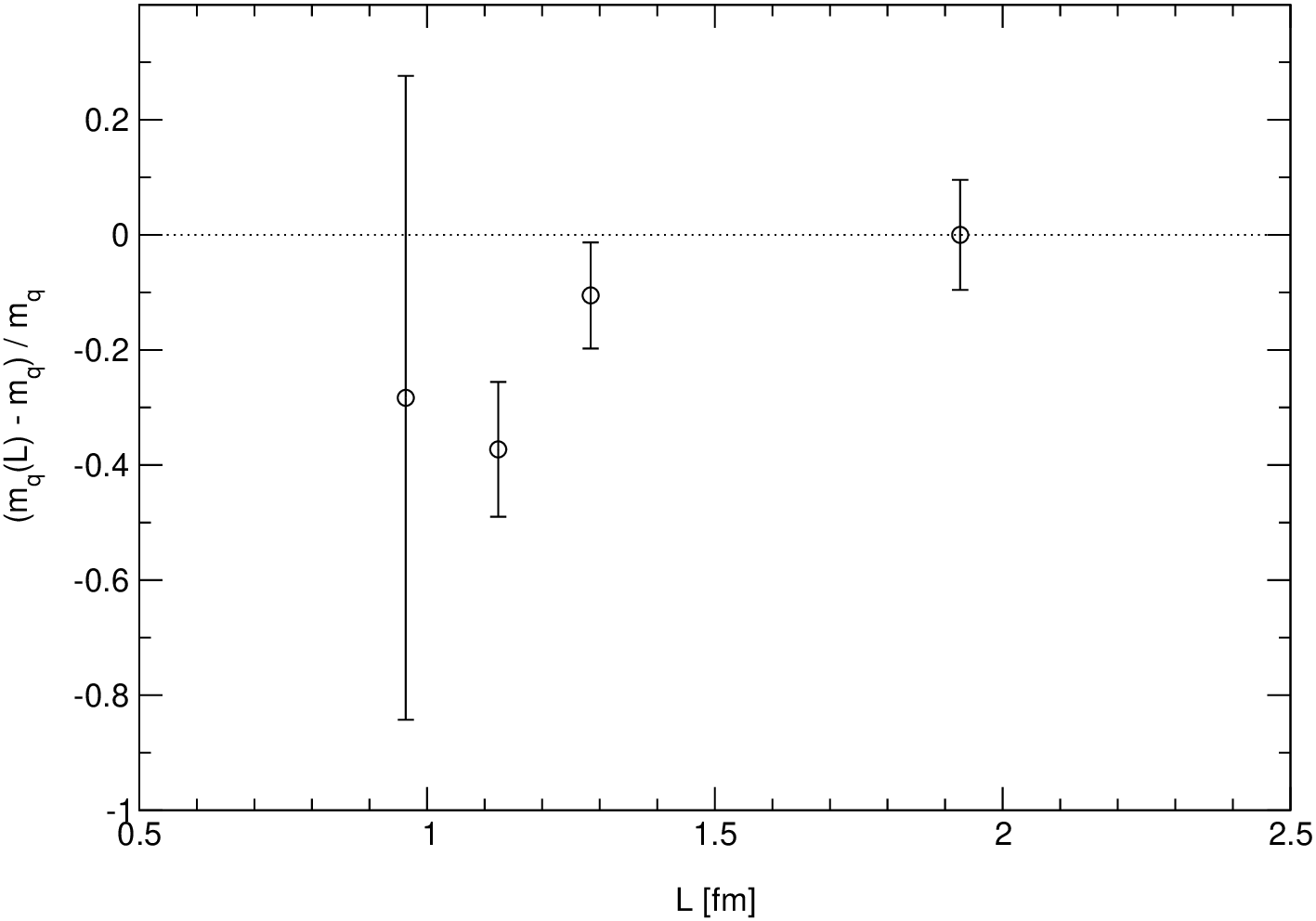}
 \caption{\label{fig:ldep_mq_5.6_1580} Same as
  Figure~\ref{fig:ldep_mq_5.6_1575} for $(\beta,\kappa)=(5.6,0.158)$.}
\end{figure}

\begin{figure}
 \includegraphics*[width=\columnwidth]{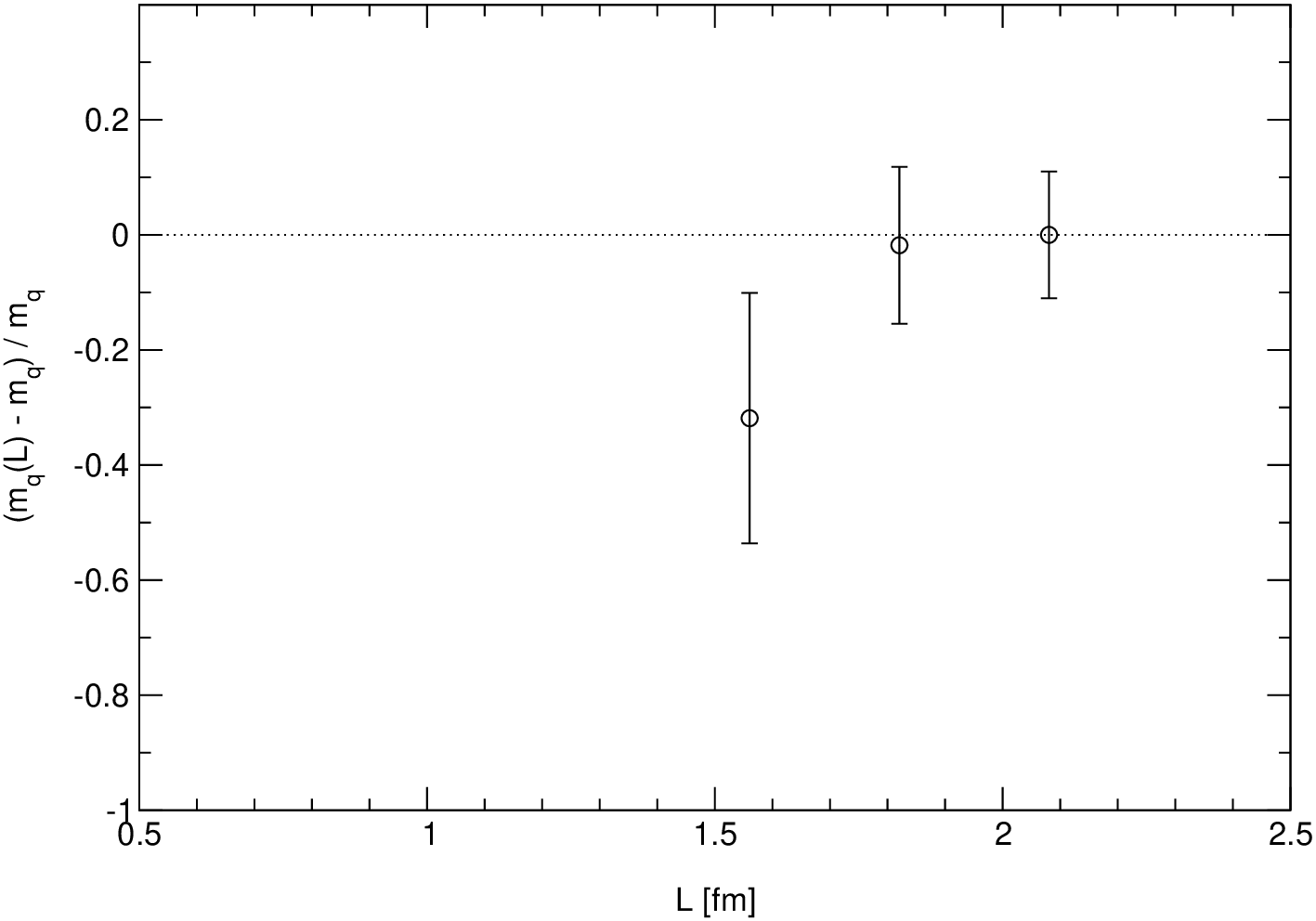}
 \caption{\label{fig:ldep_mq_5.32144_1665} Same as
  Figure~\ref{fig:ldep_mq_5.6_1575} for
  $(\beta,\kappa)=(5.32144,0.1665)$.}
\end{figure}

One method that lends itself to checking on the sig\-ni\-fi\-cance of
lattice artefacts in our simulations is to consider the PCAC relation
\begin{equation}
\label{eqn:PCAC}
\partial_\mu A_\mu^a(x) = 2m P^a(x)
\end{equation}
between the isovector axial current
\begin{equation}
A_\mu^a(x) = \qbar(x)\gamma_\mu\gamma_5 \frac{1}{2}\tau^a q(x)
\end{equation}
and the associated density
\begin{equation}
P^a(x) = \qbar(x)\gamma_5 \frac{1}{2}\tau^a q(x),
\end{equation}
where $\tau^a$ denotes a Pauli matrix acting on the flavor indices of
the quark field $q$. On the lattice, the bare quark mass $Z_q M_q$
can be extracted from ratios of correlation functions,
\begin{equation}
\label{eqn:m_PCAC}
Z_q M_q = \frac{1}{2} \frac{\ev{\partial_\mu A_\mu^a(x)
    J^a}}{\ev{P^a(x)J^a}} + O(a),
\end{equation}
where the (smeared) source $J^a$ is a suitable polynomial in the quark
and gluon fields, and $Z_q\!=\!Z_P/Z_A$. The PCAC relation
\eqref{eqn:PCAC} is an operator identity that holds for the Wilson
action up to $O(a)$ effects. Consequently, its lattice version
holds---up to those effects---for any choice of boundary conditions,
source operators and lattice sizes. This means in particular that at
fixed $\beta$ and $\kappa$ any residual lattice size dependence of the
PCAC quark mass \eqref{eqn:m_PCAC} must be a lattice
artefact. Figs.~\ref{fig:ldep_mq_5.6_1575}--\ref{fig:ldep_mq_5.32144_1665}
show the volume dependence of the relative deviation of the PCAC quark
mass $m_q(L)$ from its value $m_q$ on the largest lattices, for our
three $(\beta,\kappa)$ combinations.

We find that at $(\beta,\kappa)=(5.6,0.1575)$ the discretization errors appear
to be small for $L\gtrsim 1$~fm, while for $(\beta,\kappa)=(5.6,0.158)$ and
$(5.32144,0.1665)$ they are small only for $L\gtrsim 1.3$~fm and $L\gtrsim
1.8$~fm, respectively. On the smaller lattices the cut-off shows up in quark
mass shifts of 20--40\% (with large error bars on the smallest lattices). The
fact that the cut-off effects are small for $(\beta,\kappa)=(5.6,0.1575)$ and
somewhat larger for $(\beta,\kappa)=(5.6,0.158)$ and $(5.32144,0.1665)$ is
consistent with our observations in Section~\ref{sec:vol_dep}, where in
Fig.~\ref{fig:mNmpisq_dep} we saw no significant lattice artefacts in the
nucleon mass for $\mpi=643$~MeV, while for $\mpi=490$~MeV and $\mpi=419$~MeV
the nucleon mass displayed some deviation from the curve (which represents a
fit to $O(a)$-improved data).

\section{\label{sec:summary}Summary}

In order to investigate finite-size effects in stable light hadron masses
obtained from Lattice QCD with two dynamical flavors of Wilson fermions we
have complemented previous SESAM/\txl\ simulations at quark masses
corresponding roughly to 85 and 50\% of the strange quark mass by several runs
at the same coupling of $\beta=5.6$, but on smaller lattices. In addition we
have carried out exploratory simulations using three different lattice volumes
at a stronger coupling of $\beta=5.32144$ in an attempt to push our analysis
towards the regime of lighter quark masses. We have succeeded in simulating
near $m_s/3$, which in terms of $\mpi/\mrho$ is close to the rho decay
threshold of $0.5$. The physical extent of the investigated lattices ranges
between 0.85~fm and 2.04~fm.

We have addressed, from a practical point of view, the question to what extent
the volume dependence of the computed pion, rho and nucleon masses can be
parameterized by simple functions, and if with these functions an
extrapolation from small and intermediate lattices to the infinite volume is
possible. To this end we have compared an exponential ansatz motivated by
L\"uscher's mass-shift formula for the pion to the power law observed by
Fukugita \etal. On the basis of various fits we conclude that while the power
law may be used to describe the volume dependence of the masses at volumes
smaller than roughly $(1.5~\mathrm{fm})^3$, over the full range of simulated
lattices---and in particular with respect to the asymptotic behavior---the
exponential ansatz is more appropriate. Although the extrapolation of simple
exponential fits to the infinite volume in general provides only a lower bound
to the asymptotic mass, this bound may be close to the true asymptotic value
if the relative difference between the masses from the largest volumes
incorporated in the fit is already quite small (of the order of a few
percent). For small volumes alone, however, this is in general not the case.

The exponential parameterization corresponds in its functional form
precisely to L\"uscher's asymptotic formula for the pion (with input
from infinite-volume ChPT at leading order). Although we have found
that the single exponential allows for a reasonable phenomenological
description of our light hadron masses over a wide range of lattice
volumes, a large coefficient multiplying the exponential attests to
the fact that the data points from the small lattices lie outside the
parameter regime in which the original formula holds. In the case of
the pion we have illustrated this by a comparison of our data to
L\"uscher's formula with input from ChPT up to NNLO and to the full LO
ChPT result for the pion mass in finite volume. Of course, if an
appropriate analytic prediction with controlled errors is available it
is preferable to an extrapolation based on a fit with free
parameters. We have found, however, that in the parameter regime of
our simulations even the best currently available estimate for the
\emph{pion} finite-size effect, based on a combination of the
asymptotic L\"uscher formula with NNLO ChPT input and the full
finite-volume (but LO) ChPT result, predicts mass shifts of a few
percent only. This is comparable in size to the typical statistical
errors and therefore hard to detect in practice. Our simulations at
$\beta=5.32144$ probably are in a pion mass regime where the box-size
dependence can be described by such a formula, but more statistics is
needed to corroborate this assumption. While L\"uscher's formula with
input at next-to-leading and next-to-next-to-leading order ChPT can be
used to control the convergence of the chiral expansion, a full higher
order result from ChPT for the pion mass in finite volume would be
highly useful to fully assess the role of the sub-leading terms in the
large-$L$ expansion.

For the \emph{nucleon}, a promising finite-size mass formula has
recently become available from relativistic Baryon ChPT. We have shown
for three different pion masses (two of which are smaller than the
ones considered by the QCDSF-UKQCD collaboration) that it describes
our simulated nucleon masses remarkably well even down to box-sizes of
about 1~fm.  We have also seen that above this size it can, in
principle, be used to estimate the infinite-volume mass already on the
basis of a single measurement, provided that the corresponding
asymptotic pion mass is known. If, as in our case, data from several
lattice volumes are available, they can be combined to obtain a
reliable estimate with controllable errors.

\begin{acknowledgments}
The numerical calculations for the present work have been performed on the
Cray T3E and APE100/APEmille computers of the John von Neumann Institute for
Computing (NIC) in J\"ulich and Zeuthen, and on the Alpha-Linux cluster ALiCE
of the University of Wuppertal. We thank all these institutions for their
continuous, substantial support. We are indebted to S.~D\"urr for useful
discussions and for kindly providing us with the numerical data for the curves
in Fig.~\ref{fig:ldep_pion}. We are also grateful to R.~Sommer, I.~Montvay and
T.~R.~Hemmert for valuable hints and discussions. We thank Z.~Sroczynski for
writing the major part of our HMC code for ALiCE, and T.~D\"ussel for his help
in determining $r_0$. This work was supported by DFG grant Li701/4-1. It is
also part of the EU Integrated Infrastructure Initiative ``Hadron Physics''
(I3HP) under contract RII3-CT-2004-506078.
\end{acknowledgments}

\bibliography{bibliography,fse,lippert}

\appendix*

\section{\label{apx:parameters}Fit parameters}

\begin{table*}
 \caption{\label{tab:ldep_fit_params_56_1575}Fit parameters for
  $(\beta,\kappa)=(5.6,0.1575)$. $d=2$ for ``pow'' (Eq.~\eqref{eqn:pow}) and
  $d=1/2$ for ``exp'' (Eq.~\eqref{eqn:exp}).}
 \begin{ruledtabular}
  \begin{tabular}{lcD{,}{,}{3}llrrr}
   Fit type & \cc{$H$} & \cc{$[L_1,L_2]$} &
   \cc{$m_H\,[\mathrm{GeV}]$} & \cc{$c\,[\mathrm{GeV}^{-d}]$} & \cc{$\chisqdf$} &
   \cc{$\Delta(L_\mathrm{max})$} & \cc{$\Delta(L\!=\!\infty)$} \\
   \hline
   pow & PS & 10,24 & 0.570(35) & \phantom{1}41.2(7.1) & 31.65 & -5.57\% & -11.40\% \\
   pow & V  & 10,24 & 0.872(19) & \phantom{1}35.1(5.3) & 3.34 & -1.38\% & -4.87\% \\
   pow & N  & 10,24 & 1.255(43) & \phantom{1}88.5(9.3) & 5.83 & -2.99\% & -8.84\% \\
   exp & PS & 10,24 & 0.624(13) & \phantom{1}65.9(4.2) & 5.59 & -2.41\% & -2.91\% \\
   exp & V  & 10,24 & 0.9125(92) & \phantom{1}63.5(5.6) & 1.19 & -0.17\% & -0.51\% \\
   exp & N  & 10,24 & 1.372(22) & 142.7(9.5) & 2.37 & 0.18\% & -0.32\% \\
  \end{tabular}
 \end{ruledtabular}
\vspace{2ex}
 \caption{\label{tab:ldep_fit_params_56_1580}Same as
  Table~\ref{tab:ldep_fit_params_56_1575} for $(\beta,\kappa)=(5.6,0.158)$.}
 \begin{ruledtabular}
  \begin{tabular}{lcD{,}{,}{3}llrrr}
   Fit type & \cc{$H$} & \cc{$[L_1,L_2]$} &
   \cc{$m_H\,[\mathrm{GeV}]$} & \cc{$c\,[\mathrm{GeV}^{-d}]$} & \cc{$\chisqdf$} &
   \cc{$\Delta(L_\mathrm{max})$} & \cc{$\Delta(L\!=\!\infty)$} \\
   \hline
   pow & PS & 12,24 & 0.417(33) & \phantom{1}55.3(9.2) & 8.55 & -2.67\% & -14.81\% \\
   pow & V & 12,24 & 0.780(59) & \phantom{1}62.5(13.1) & 5.22 & -2.10\% & -9.86\% \\
   pow & N & 12,24 & 1.0894(92) & 124.0(2.3) & 0.07 & -0.44\% & -11.30\% \\
   exp & PS & 12,24 & 0.466(20) & \phantom{1}47.9(4.2) & 3.86 & -1.50\% & -4.91\% \\
   exp & V & 12,24 & 0.836(45) & \phantom{1}53.1(10.0) & 4.31 & -1.27\% & -3.41\% \\
   exp & N & 12,24 & 1.208(20) & 104.1(5.1) & 0.50 & 1.30\% & -1.65\% \\
  \end{tabular}
 \end{ruledtabular}
\vspace{2ex}
 \caption{\label{tab:ldep_fit_params_532144_1665}Same as
  Table~\ref{tab:ldep_fit_params_56_1575} for
  $(\beta,\kappa)=(5.32144,0.1665)$.}
 \begin{ruledtabular}
  \begin{tabular}{lcD{,}{,}{3}llrrr}
   Fit type & \cc{$H$} & \cc{$[L_1,L_2]$} &
   \cc{$m_H\,[\mathrm{GeV}]$} & \cc{$c\,[\mathrm{GeV}^{-d}]$} & \cc{$\chisqdf$} &
   \cc{$\Delta(L_\mathrm{max})$} & \cc{$\Delta(L\!=\!\infty)$} \\
   \hline
   pow & PS & 12,16 & 0.428(22) & -15.0(16.3) & 2.09 & -0.83\% & 2.23\% \\
   pow & V & 12,16 & 0.743(32) & \phantom{-}23.3(29.0) & 0.77 & 0.54\% & -2.08\% \\
   pow & N & 12,16 & 1.037(63) & \phantom{-}90.9(52.8) & 1.87 & 0.98\% & -6.05\% \\
   exp {\scriptsize ($c\!=\!C$)} & PS & 12,16 & 0.4089(81) & &2.04 & -2.36\% & -2.36\% \\
   exp & V & 12,16 & 0.756(20) & \phantom{-}18.4(26.7) & 0.86 & 0.64\% & -0.31\% \\
   exp & N & 12,16 & 1.088(42) & \phantom{-}74.9(50.5) & 2.32 & 1.19\% & -1.46\% \\
  \end{tabular}
 \end{ruledtabular}
\end{table*}

\end{document}